\documentclass{elsarticle}
\usepackage[normalem]{ulem}
\usepackage{amsmath}
\usepackage{epstopdf}
\usepackage{flushend}
\usepackage{indentfirst}
\usepackage{amssymb}
\usepackage[figuresright]{rotating}
\usepackage{caption}
\usepackage{subfigure}
\usepackage{bm}
\usepackage{natbib}
\usepackage[dvipsnames]{xcolor}
\usepackage{multirow}

\begin{document}

\begin{frontmatter}

\title{Development of a thermo-mechanically coupled crystal plasticity
modeling framework: application to polycrystalline homogenization}

\author[add1,add2]{Jifeng Li}
\author[add1,add3]{Ignacio Romero}
\author[add1,add2]{Javier Segurado\corref{cor1}}
\address[add1]{IMDEA Materials Institute, Eric Kandel 2, Getafe, 28906 Madrid, Spain}
\address[add2]{E.T.S. de Ingenieros de Caminos, Universidad  Polit{\'e}cnica de Madrid, 28040 Madrid, Spain}
\address[add3]{ETSI Industriales, Jose Gutierrez Abascal 2, Universidad Polit{\'e}cnica de Madrid, 28006 Madrid, Spain}
\cortext[cor1]{Corresponding author. E-mail: javier.segurado@imdea.org(Javier Segurado)}

\begin{abstract} Accurate predictions of thermo-mechanically coupled process in metals can lead to a
reduction of cost and an increase of productivity in manufacturing processes such as forming. For modeling these coupled processes with the finite element method, accurate descriptions of both the
mechanical and the thermal responses of the material, as well as their interaction, are
needed. Conventional material modeling employs empirical macroscopic constitutive relations but   does
not account for the actual thermo-mechanical mechanisms occurring at the microscopic level.
However, the consideration of the latter might be crucial to obtain accurate predictions and a
complete understanding of the underlying physics.

In this work we describe a fully coupled implicit thermo-mechanical framework for crystal plasticity
simulations. This framework includes thermal strains, temperature dependency of the crystal behavior
and heat generation by dissipation due to plastic slip and allows the use of large deformation steps thanks to the
implicit integration of the governing equations. Its use within computational homogenization
simulations allows to bridge the plastic deformation and temperature gradients at the macroscopic
scale with the microscopic slip at the grain scale. A series of numerical examples are presented to
validate the approach.

\end{abstract}


\end{frontmatter}

\section{Introduction}

Metallic materials usually undergo large plastic deformation under a wide range of strain rates and temperatures during
forming and other manufacturing processes.  Accurate predictions of the material behavior in these situations are of
great interest for industry since they are helpful for reducing the manufacturing cost and increasing the productivity.
The response of metals in these processes is, however, very complex since their mechanical behavior is often strongly
coupled with thermal phenomena. For example, the thermal variations may produce non-negligible effects on the mechanical
behavior as the result of thermal expansion and/or contraction. Also, temperature changes may modify significantly the
mechanical behavior of a material introducing softening, increasing its rate sensitivity, and may even cause phase
transformations. On the other hand, the mechanical processes influence the thermal field via heat generation
caused by plastic work dissipation, as well as through changes in the thermal boundary conditions due to the geometrical
evolution of the material configuration. Thus, accurate simulations of those thermo-mechanical processes require
a fully coupled modeling framework that can describe with precision both the mechanical and the thermal behaviors of
the material, as well as their interaction.

The solution of a thermo-mechanical problem involves the consideration of the balance of momentum and balance of energy equations. Both of them are standard for small and large strain solid mechanics, but the material response to mechanical and thermal loading is very complex,
more so if their coupling is  considered. Given those difficulties, the thermo-mechanical response of metals is often modeled with phenomenological models such as Johnson-Cook's \cite{Johnson1983}, Zerilli-Armstrong \cite{Zerilli19871816}, and the Mechanical Threshold stress model \cite{FOLLANSBEE198881}. These constitutive relations, and others employed in simulations, consciously ignore microscopic phenomena that take place at the sub-grain level, even when it is widely acknowledged that are precisely these which determine the mechanical behavior.

In order to develop models that can provide a higher degree of accuracy in their predictions it is mandatory to incorporate
into them information regarding the microscopic behavior of the material. In particular, it has been observed experimentally that polycrystalline metals exhibit a marked anisotropic response during plastic deformation, and it is mainly caused by the crystallographic texture resulting from the reorientation of the grains. Hence, incorporating the evolution of texture in a detailed constitutive relation for the thermo-mechanical behavior of metallic polycrystals is absolutely required.

To this end, crystal plasticity (CP) was developed in the late 70s and 80s
\cite{HILL196695,RICE1971433,ASARO1977309,PEIRCE19831951} as a constitutive equation into the framework of
continuum mechanics, based on the model envisioned by Taylor in the 30s.  CP models predict the
mechanical behavior of a single crystalline metal taking into account the dislocation slip through
the slip planes characteristic of its lattice. The use of CP models within homogenization frameworks
allows to simulate the response of a full polycrystal including texture evolution by explicitly
considering grain orientation and shape changes in the micro-scale (see \cite{Segurado2018} for a
review). Such type of models, due to their verified predictive capacity, have been used in the prediction of the mechanical response under monotonic loading \cite{CRUZADO2015242,ZLE15,HERRERASOLAZ20141,ZHOU201619,SG16,GST16,Rui2017} , cyclic loading \cite{LUO2013,MUHAMMAD2017137,CRUZADO2017148} and  fatigue response
\cite{DUNNE20071061,BRIDIER20091066,ROVINELLI2018208,CRUZADO201840,CHEN2018213} or to simulate forming processes such as rolling \cite{SARMA200491,WR07,MBB15, GUPTA2018168}.

Temperature effects on the mechanical response of a polycrystal using computational models with
realistic microstructures have been introduced in many studies
\cite{BT08,CHEONG20051797,Rodriguez2015191,HU20161} using a standard, purely mechanical framework,
neglecting the effect of thermal strains and not considering heat generation by plastic
dissipation. The structural heating due to plastic dissipation in crystal plasticity
has been accounted for as responsible for the \emph{local} temperature increase using a Taylor model
to homogenize the polycrystal response  \cite{KOTHARI199851,HAKANSSON20081570}. Under the Taylor
approach, individual grain orientations are considered, therefore accounting for the effect of
texture. On the other hand, the state of each grain is represented by a single value for each field
variable and all the crystals share the same deformation gradient, limiting the
accuracy of the models. Under this constraint, heat conduction at the microscale is neglected and
instantaneous temperature equilibrium is assumed. More recently, computational polycrystalline
homogenization based on finite elements has been used to model the thermo-mechanical response of a
polycrystal considering a detailed representation of the microstructure and resolving the
microfields within the grains. Most of the studies \cite{ROSSITER20101702,BARGMANN2013899} use
microscopic adiabatic conditions and therefore neglect thermal transport phenomena at the
microscale, resulting in a very heterogeneous temperature distribution due to the differences in
plastic dissipation in different points of the microstructure. In the last years, two fully coupled
models for crystal plasticity have been presented that include heat diffusion at the microscale. In
\cite{SEGHIR20101155}, a fully coupled thermodynamical framework using implicit integration is
developed. The model includes heat generation and thermal strains but the crystal plastic behavior
was assumed to be independent of the temperature. An alternative thermo-mechanical coupled framework
was proposed recently for FCC materials \cite{CYR2015166} including a more physical description of
the plastic flow in the crystal plasticity model, but formulated using an explicit time integration.
It must noted, finally, that none of the works reviewed proposing coupled thermo-mechanical
frameworks for polycrystals considers the use of periodic boundary conditions for both thermal and
mechanical fields, limiting the accuracy of the schemes proposed due to the inaccuracies near the
RVE surfaces.

In this work we propose a fully coupled thermo-mechanical framework for computational homogenization of polycrystals  to bridge the plastic deformation and temperature gradients at the macroscopic scale with the microscopic slip at the grain scale. This framework includes (1) Strong thermo-mechanical coupling, including thermal strains and temperature dependency of the crystal behavior through a physically based crystal plasticity model including temperature dependency of plastic slip and heat generation by dissipation due to plastic deformation. (2) Periodic boundary conditions in both the mechanical and thermal problems to account in the most accurate way with the polycrystalline homogenization problem or FE2 procedure. (3) An implicit implementation of the equations. (4) An experimental benchmark in a real polycrystal including several cases to asses the validity and accuracy of the model.

The thermo-mechanical homogenization problem is solved using Finite Element simulations in
Representative Volume Elements of the microstructure (RVEs) and periodic boundary conditions.
During the deformation heat is generated at the grain scale due to plastic dissipation and then
diffused throughout the body giving rise to thermal deformations and, thanks to the use of a
temperature dependent CP formulation, modifying the mechanical response of the crystals. Both the
crystal plasticity model and the heat conduction model are implemented in the commercial finite
element code ABAQUS/Standard.  In addition to the computation polycrystalline framework,
the results of a homogenization model based on the Taylor approach are presented to show the effect of
considering heat conduction at the microscale.

The remainder of this paper is organized as follows. In Section 2, a brief description of a general thermo-mechanical
coupled problem is introduced, followed by a detailed introduction of the crystal plasticity model for single crystal
and the computational homogenization approach that extracts the macroscopic material behavior from the finite element
solution of a boundary value problem of a representative volume element. In Section 3, the numerical method for updating the deformation field and the temperature field is presented. In Section 4, a series of numerical examples are presented to validate the proposed framework. Finally, some concluding remarks are drawn.

\section{Thermo-Mechanically Coupled Crystal Plasticity Modeling Framework}

In this section, we start from the discussion of the general thermo-mechanical coupled analysis,
and then we give a detailed introduction of the crystal plasticity
constitutive model and the two homogenization approaches considered,  the Taylor model and the full field computational homogenization approach.

\subsection{Thermo-Mechanical Coupled Analysis}

A thermo-mechanical problem is governed by the balance
equations of linear momentum and energy. For quasi-static
problems, the former can be written as
\begin{equation}
  \nabla \cdot \bm{\sigma} + \bm{b} = \bm{0} ,
  \label{eq-balance-p}
\end{equation}
where $\bm{\sigma}$ is the Cauchy stress tensor, $\bm{b}$ is the body force per unit of deformed volume, and $\nabla\cdot$ is the spatial divergence operator.
This equilibrium equation must hold for every $\bm{x}=\bm{\varphi}(\bm{X},t)$,
where $\bm{X}$ denotes the coordinates of points of the body $\mathcal{B}_0$ in the
reference configuration, $t$ refers to time, and
$\bm{\varphi}$ is the deformation mapping. The boundary $\partial\mathcal{B}_0$ can
be split in two disjoint parts denoted $\partial_\varphi\mathcal{B}_0$ and
$\partial_t\mathcal{B}_0$. Then, in addition to equilibrium condition~\eqref{eq-balance-p} of
points inside $\mathcal{B}_0$, the deformation must satisfy $\bm{\varphi}=\bar{\bm{\varphi}}$
on $\partial_\varphi\mathcal{B}_0$ and $\bm{\sigma}\bm{n} = \bm{t}$ on $\partial_t\mathcal{B}_0$,
where $\bm{n}$ is the outer normal to the boundary and $\bm{t}$ is a known
field of surface tractions.

The second equation that describes the thermo-mechanical problem
is the one imposing the balance of energy. If
there are no external heat sources, and heat conduction
follows Fourier's law, this equation reads
\begin{equation}
  \rho c \dot{\theta}
  =
  \nabla\cdot(\bm{k} \nabla \theta) + \dot{Q}_V  .
  \label{eq-balance-e}
\end{equation}
Here $\theta$ denotes the absolute temperature,
$\rho$ refers to the mass per unit of deformed volume,
$c$ is the specific heat capacity,
$\bm{k}$ is the spatial conductivity tensor,
and $\dot{Q}_V$ is the internal heat generation rate
per unit of deformed volume. Eq.~\eqref{eq-balance-e}
must be satisfied in every point of the interior $\mathcal{B}_0$.
As in the case of the mechanical problem, the boundary
of the body can be split into two disjoint sets $\partial_h\mathcal{B}_0$
and $\partial_\theta\mathcal{B}_0$. The thermal problem
is completed with Dirichlet boundary conditions for
the temperature on $\partial_\theta\mathcal{B}_0$ and
Neumann conditions for the heat flux on $\partial_h\mathcal{B}_0$.

The evaluation of the stress tensor $\bm{\sigma}$, as a function
of the current and past values of the deformation and temperature,
is the central problem of solid thermo-mechanics. In continuum
thermo-mechanics with internal variables, it is assumed that
the stress tensor depends on the deformation gradient $\bm{F}
=\nabla_0\bm{\varphi}$, its rate $\dot{\bm{F}}$, the local
value of $\theta$ and, possibly,
some internal variables denoted collectively as $\bm{\xi}$. That is,
\begin{equation}
  \bm{\sigma}
  =
  \hat{\bm{\sigma}}
  \left( \bm{F}, \ \dot{\bm{F}}, \ \theta, \ \bm{\xi} \right)  .
  \label{eq-constitutive}
\end{equation}
In the definition of $\bm{F}$ and below, $\nabla_0$ refers
to the material gradient operator.

The constitutive law described by Eq.~\eqref{eq-constitutive}
strongly couples Eqs.~\eqref{eq-balance-p} and~\eqref{eq-balance-e},
and hence the deformation and temperature fields. As will be shown
later when a specific form of this constitutive relation is
provided, temperature changes induce thermal strains and
changes in the mechanical constants; in turn, the dissipation
due to the power of the stress conjugated with the viscoplastic
strains generates internal heat $\dot{Q}_V$.  efficiency of the conversion of mechanical into thermal energy is not perfect because some energy is stored in microstructural transformations at the dislocation level. However, considering this phenomena from a physical view point implies the introduction of constitutive hypotheses that are material dependent (see, e.g., the discrete dislocation model presented by Benzerga et. al. for copper \cite{BENZERGA20054765}). A general framework should rely on the experimental measure of the fraction of energy transformed in heat through a coefficient that might vary from case to case. This type of approach was proposed by Taylor and Quinney by introducing a coefficient expressing the efficiency of this transformation. This is the approach taken here,  more specifically, the internal heat generation is customarily expressed as
\begin{equation}
  \label{heatGen}
  \dot{Q}_V = \chi \dot{W}^p
\end{equation}
where $\dot{W}^p$ is the plastic power and $\chi$
is the so-called Taylor-Quinney parameter, usually
selected in the range $0.85 \le \chi \le 1.0$.

As a result of the previous arguments, we conclude that an accurate
solution of both the mechanical and thermal problems is required in
order to solve the fully coupled problem. Moreover, the constitutive
relation is identified as the critical and more complex part of
the solution, and responsible for the coupling. Based on this
reasoning, and in order to get the most reliable possible model
for metallic materials, a general crystal plasticity model is chosen
in which plastic flow is aided by thermal activation. This
type of temperature dependency of the plastic behavior can be found in BCC
metals, in HCP deforming under pyramidal slip or in FCC alloys hardened by precipitates.

\subsection{Crystal Plasticity Model for a Single Crystal}
\label{subs-cp}

For finite strain inelastic mechanics of single crystal plasticity the deformation gradient $\bm{F}$
is assumed to be multiplicatively decomposed in its elastic and plastic parts  as in
$\bm{F}=\bm{F}^e\bm{F}^p$ \cite{Lee196719}.  The elastic and plastic deformation gradients  define two
different \emph{local} deformed configurations: one \emph{intermediate} configuration described
by $\bm{F}^p$, and the final deformed configuration locally determined by $\bm{F}^e\bm{F}^p$. For a
thermo-mechanical problem, a third configuration is introduced accounting for thermal deformations
\cite{MEISSONNIER2001601}. The resulting decomposition reads
  \begin{equation} \label{defGrdDcmp}
    \bm{F}=\bm{F}^e \bm{F}^p \bm{F}^{\theta}, \quad \mbox{det}\, \bm{F}^e > 0, \quad \mbox{det}\,
    \bm{F}^p = 1, \quad \mbox{det}\, \bm{F}^\theta > 0 .
  \end{equation}
where $\bm{F}^\theta$ now represents the thermal part of the deformation
gradient. The elastic deformation gradient embodies the
elastic distortions of the lattice, and the rigid body motions;
the plastic part, $\bm{F}^p$, describes the irreversible
deformations of the lattice, associated with plastic shearing along
crystallographic planes; finally, the thermal contribution $\bm{F}^\theta$
includes the lattice distortion due solely to thermal effects.
The local configuration defined by $\bm{F}^p \bm{F}^{\theta}$ is hereafter called the intermediate
configuration, and can be regarded as being obtained by a
pure elastic unloading from the current configuration.

The constitutive model relates the three parts of the deformation gradient with
the stress and temperature, or their evolution. In particular, the evolution
of the thermal deformation gradient of any metal can be expressed respect its lattice symmetry axis
as
\begin{equation}
  \label{evolveFTheta}
  \dot{\bm{F}}^\theta {\bm{F}^\theta}^{-1} = \dot{\theta} \bm{\beta}  ,
\end{equation}
where $\bm{\beta}$ is a diagonal tensor of anisotropic thermal expansion coefficients
\begin{equation}
  \bm{\beta}
  =
  \mbox{diag} \left(\beta_1,\ \beta_2,\ \beta_3 \right) .
\end{equation}

Assuming that the plastic deformation takes place purely through  dislocation glide, the evolution
of the plastic deformation gradient could be expressed as a combination of the slip rate
$\dot{\gamma}^\alpha$ in every slip system $\alpha$ as in
\begin{equation}
  \label{Lp}
  \dot{\bm{F}}^p {\bm{F}^p}^{-1}
  =
  \sum_{\alpha} \dot{\gamma}^{\alpha} \bm{S}^{\alpha}_0,
\end{equation}
where, for every slip system $\alpha$,
$\bm{S}^{\alpha}_0=\bm{s}^{\alpha}_0 \otimes \bm{m}^{\alpha}_0$
is the Schmidt tensor, $\bm{s}^{\alpha}_0$ and $\bm{m}^{\alpha}_0$ are unit vectors
along the slip direction and normal to the slip plane, respectively.

In the intermediate configuration, the elastic Green-Lagrange strain
$\bm{E}^e$ is defined as
\begin{equation}
  \label{GreenStrain}
  \bm{E}^e = \frac{1}{2}\left({\bm{F}^e}^T{\bm{F}}^e - \bm{I} \right),
\end{equation}
where $\bm{I}$ is the second order identity tensor.
Let $\bm{S}^*$ be second Piola-Kirchhoff tensor, the stress that is
work conjugate to $\bm{E}^e$.
An elastic constitutive relation is given by
\begin{equation} \label{PK2}
  \bm{S}^*
  = \mathcal{L} : \bm{E}^e ,
\end{equation}
where $\mathcal{L}$ is the fourth-order temperature dependent elasticity tensor.

The Cauchy stress is the push forward of the stress $\bm{S}^*$
to the current configuration, that is
\begin{equation}
  \label{CauchyStress}
  \bm{\sigma} =
  \left(\mbox{det}\, \bm{F}^e \right)^{-1} \bm{F}^e \bm{S}^* {\bm{F}^e}^T.
\end{equation}

To close the mechanical constitutive law, the evolution of the plastic
slip $\gamma^{\alpha}$ must be specified for every slip plane. In
a viscoplastic model, the slip rate is expressed with the flow model
\begin{equation}
  \dot{\gamma}^\alpha
  =
  \hat{\dot{\gamma}}^\alpha \left(\tau^\alpha, s^\alpha, \theta \right) ,
\end{equation}
where $\tau^\alpha$ is the resolved shear stress in slip system $\alpha$, defined
as
\begin{equation} 
  \tau^\alpha = \mbox{det}\, \bm{F}^\theta
  \left( {\bm{F}^\theta}^T \bm{S}^* {\bm{F}^\theta}^{-T} \right) \cdot \bm{S}^\alpha_0,
  \end{equation}
and $s^\alpha$ is the slip resistance  \cite{MEISSONNIER2001601}. In this
equation, and below, the dot product refers to the full contraction of vectors
or tensors over all their indices.  The quantity $s^\alpha$ evolves following
the hardening model
\begin{equation}
  \dot{s}^{\alpha} = \sum_{\beta}{h^{\alpha\beta}\left| \dot{\gamma}^{\beta}\right|} ,
\end{equation}
where
\begin{equation}
  \label{hAlpha} h^{\alpha\beta}=h^{\alpha}q^{\alpha\beta}
\end{equation}
represents the hardening modulus, $q^{\alpha\beta}$ stands for
the latent hardening coefficients, and $h^{\alpha}$ is the self hardening modulus.

Several flow and hardening models have been proposed
in the literature \cite{ASARO1985923, Frost1982, PEIRCE19821087, BROWN198995}. In this work, a physically based
flow model including temperature dependency is combined with a phenomenological hardening model in order to provide a
relatively general CP framework that can be adapted in principle to many different alloys \cite{Rodriguez2015191,KOTHARI199851}.

Assume that the slip resistance $s^\alpha$ could be decomposed as
\begin{equation}
  s^\alpha = s^\alpha_{at} + s^\alpha_t ,
\end{equation}
where $s_t$ and $s_{at}$ denote the thermal and athermal parts due to thermally activated obstacles and athermal obstacles,
respectively \cite{KOTHARI199851}. The resolved shear stress at slip system $\alpha$ needs to overcome the athermal part of the slip resistance
before slip starts. Beyond that value of the resolved shear stress, the thermal energy required to overcome the obstacle
depends on the effective resolved shear stress $  \tau^{\alpha}_{eff}$, defined as
\begin{equation}
  \tau^{\alpha}_{eff} = \left| \tau^{\alpha} \right| - s^{\alpha}_{at} .
\end{equation}
According to the flow model proposed by Frost and Ashby\cite{Frost1982}, the slip rate in a slip system $\alpha$ controlled by some thermal
activation process can be obtained from the Orowan equation
  \begin{equation}
    \dot{\gamma}^\alpha = \rho^\alpha_m b^\alpha \bar{L} \bar{\nu}^\alpha(\tau^{\alpha}_{eff},\theta).
  \end{equation}
where $\rho^\alpha_m$ denotes the mobile dislocation density, $b^\alpha$ the Burgers vector of that system and $\bar{L}$
is a characteristic length of the distance between obstacles. $\bar{\nu}^\alpha$ is the average frequency of jumping events
across the obstacles, depends both on the effective resolved stress and the current temperature and is obtained using the theory
of thermally activated processes. If all the constants in the previous equation are grouped, a single parameter independent
of the temperature $\dot{\gamma}^\alpha_0$ , the reference slip rate, can be defined as
\begin{equation}
\dot{\gamma}^\alpha_0 \approx \rho^\alpha_m b^\alpha \bar{L} \nu_0,
\end{equation}
where $\nu_0$ is the Debye frequency. Using this value, the shear rate can finally be expressed as
\begin{equation}
  \dot{\gamma}^{\alpha} =
  \left\{
    \begin{aligned}
      &0, \  &\tau ^{\alpha}_{eff} \le 0,  \\
      &\dot{\gamma}^\alpha_0 \mbox{exp}\left(-\frac{\Delta
          G^{\alpha}}{k_B\theta}\right)\mbox{sign}\left(\tau^{\alpha}\right), \
      &0<\tau ^{\alpha}_{eff} \le s ^{\alpha}_{t},  \\
  \end{aligned}
  \right.
  \end{equation}
where $k_B$ is Boltzmann's constant and $\Delta G^{\alpha}$ is the activation enthalpy. The latter
can be obtained with the expression
\begin{equation} 
  \Delta G^{\alpha} =
  \Delta F^{\alpha} \left[ 1 - \left(\frac{|\tau^{\alpha}_{eff}|}{s^{\alpha}_t}\right)^p \right]^q,
\end{equation}
with $\Delta F^{\alpha}$ being the activation energy at 0 K, $p$ and $q$ denoting
model parameters with values $0<p<1$ and $1<q<2$.

The flow model based on thermal activation is valid for any alloy in which dislocation glide is controlled by thermal
activation. In BCC crystals, the lattice friction is the main energy barrier and it is usually assumed that the thermal
part of the slip resistance $s^\alpha_t$ is constant \cite{KOTHARI199851}. For FCC crystals, the short range
dislocation-dislocation interactions are the main thermal process and it was investigated experimentally
\cite{Cottrell195517} that both the thermal and athermal parts of the slip resistance evolve with
the deformation but their ratio $\left. s^{\alpha}_t \middle / s^{\alpha}_{at} \right.$
could be regarded as a constant over a broad range of strains. In this study $s^\alpha_t$ is considered independent of
the accumulated plastic slip (BCC case), and the self hardening modulus $h^\alpha$ in Eq.~(\ref{hAlpha}) for describing
the evolution of the slip resistance is given by a phenomenological hardening model as
\begin{equation}
  h^{\alpha}
  = h^{\alpha}_0 {\left( 1-s^{\alpha}_{at}/s^{\alpha}_s \right)}^{r_1}\mbox{sign}
  \left( 1-s^{\alpha}_{at}/s^{\alpha}_s\right).
\end{equation}
where $h^{\alpha}_0$ is the initial hardening moduli, $r_1$ is a model parameter,
and $s^{\alpha}_s$ is the saturation value of $s^{\alpha}_{at}$.

Finally, the volumetric heat generation contributed from plastic work dissipation defined in
Eq.~\eqref{heatGen} can be evaluated to be
\begin{equation}
\dot{Q}_V = \chi \sum_{\alpha} \tau_\alpha \dot{\gamma}_\alpha .
\label{taylorquinney_CP}
\end{equation}

\subsection{Polycrystal computational homogenization}
\label{subs-homo}

For metallic materials, every macroscopic material point is associated with a collection of single crystals characterized
by their shapes and orientations, and the behavior of a material point is determined by the collective responses of the
single crystals or grains. The relation between the macroscopic or polycrystalline response and the crystal behavior depends on the actual microstructure (grain shape and orientation distributions) and is given by the theory of homogenization. The simplest approach is the so-called Taylor model \cite{Taylor1938307}, which assumes that the local deformation gradient in each grain is constant and identical to the deformation gradient of the macroscopic material point. In this case each grain is considered to be an independent single crystal with prescribed initial volume and orientation, while interactions between neighboring grains are neglected. The macroscopic response of the material point is
extracted simply by averaging the responses of all the associated grains. More refined homogenization models also consider a constant value of the fields in each grain (mean-field models) but allow that each grain deforms independently based on its actual orientation. The viscoplastic self-consistent model (VPSC) is the most common approach within mean-field approaches \cite{MOLINARI19872983, LEBENSOHN19932611, Tome2005} and due to its accuracy and relative simplicity can be used as constitutive equation of a polycrystal in macroscopic simulations \cite{SEGURADO2012124}. However, meanfield approaches are not accurate enough for some purposes and computational homogenization schemes are preferred.

The computational homogenization of polycrystals is based on the numerical resolution, usually with the Finite Element method, of the governing equations at the microscopic level in a representative volume element
(RVE) of the polycrystalline microstructure \cite{Roters2010,Segurado2018}. The RVE typically contains hundreds of grains, each one with an assigned initial orientation and represented by a subset of material points so that the orientation distribution function (ODF) and the distribution of the grain size in the RVE reproduce, in a statistical sense, the corresponding distribution functions experimentally measured.  The macroscopic fields are obtained as the volume average of the microscopic fields obtained by the numerical resolution of a boundary value problem in the RVE. Compared to mean-field approaches, computational homogenization accounts for a more detailed description of the microstructure evolution and the full resolution of the fields at the micro-scale result in a more accurate prediction of the macroscopic behavior of the material.

Computational homogenization is the basis of the model proposed in this work to
describe the thermo-mechanical response of a polycrystal. Nevertheless, a simpler homogenization
model based on Taylor's approach is also considered in order to show the effect in the mechanical
response of considering heat diffusion at the microscale. The main equations describing both
homogenization approaches are given next.

  \subsubsection*{Taylor model}

The Taylor model is a mean-field approach and assumes that the deformation gradient in each grain is
represented by its mean value and is identical to the macroscopic deformation gradient,
$\bar{\bm{F}}$.  In the thermo-mechanical case, the temperature in each grain is also represented by its mean value $\theta^{(k)}$ and it is considered that each grain only contributes to its local temperature rise, assuming an adiabatic condition at the microscale. Under these approaches, micro- and macro- fields are related as

\begin{equation}\label{eq:Taylor}
\begin{split}
 \bar{\bm{F}} &=
  {\bm{F}}^{(k)} \\
 \bar{\bm{\sigma}} &=
 \frac{1}{N_c} \sum^{N_c}_{k=1} {\bm{\sigma}}^{(k)} \\
 \bar{\theta}  &=
\frac{1}{N_c} \sum^{N_c}_{k=1} {\theta}^{(k)}
\end{split}
\end{equation}

where $N_c$ is the total number or grains/orientations considered to represent the polycrystal,
$\bar{\bm{F}}$, $\bar{\bm{\sigma}}$ and $\bar{\theta}$ stand, respectively,
for the macroscopic values of the deformation gradient, stress and temperature and $k$ labels a grain.

\subsubsection*{Computational homogenization framework}
Our computational homogenization framework is based on solving
  numerically the governing Eqs.~\eqref{eq-balance-p} and~\eqref{eq-balance-e}
  with periodic boundary conditions using the Finite Element method.
The stress tensor and the temperature of a macroscopic material point are calculated, respectively, with the averages:
  \begin{equation}
  \begin{split}
   \bar{\bm{\sigma}}
   &= \frac{1}{V} \int{\bm{\sigma}} \,\mathrm{d} V
   =
   \frac{1}{N} \sum^{N}_{k=1} {\bm{\sigma}}^{(k)} W^{(k)}
   \\
   \bar{\theta}
   &=
   \frac{1}{V} \int{\theta} dV
   = \frac{1}{N} \sum^{N}_{k=1} {\theta}^{(k)} W^{(k)},
  \end{split}
  \end{equation}
where $N$ is the total number of integration points within the RVE
and $W^{(k)}$ is the volume element associated to the $k$-th quadrature point.

The microstructure of a polycrystalline material can be idealized as an infinite periodic arrangement of a RVEs. Under
this approach, periodic boundary conditions need to be imposed to enforce shape compatibility between adjacent deformed
RVEs. Periodic boundary conditions reproduce exactly the deformation of an infinite media constructed by a periodic
arrangement of the RVE. Moreover, it has been observed that, among the different boundary conditions that fulfill the Hill-Mandel condition, periodic boundary conditions give solutions that converge faster than any other to the solution of an infinite RVE  \cite{Bohm}.
For a cubic RVE, let $P^+_i$ be an arbitrary node lying on a boundary surface $S^+_i$, and $P^-_i$
be its counterpart lying on surface $S^-_i$, where $S^+_i$ and $S^-_i$ are a pair of surfaces perpendicular to the $X_i$ axis.
According to the requirement of periodicity and continuity of boundary conditions, the deformations of nodes
$P^+_i$ and $P^-_i$ are linked by the multiple point constraints given by \cite{SEGURADO20133},
\begin{equation}\label{pbcDisp}
  \bm{u}(P^+_i) - \bm{u}(P^-_i) = (\bar{\bm{F}} - \bm{I})\bm{L}_i  = \bm{u}(M_i) - \bm{u}(O)
  \quad (i = 1,2, 3),
\end{equation}
where $\bar{\bm{F}}$ is the deformation gradient of the corresponding macroscopic material point, $\bm{L}_i$ is a
vector with the length of the cubic RVE along the $X_i$ axis and $O$ and $M_i$ being its two associated vertices.
The vertex $O$ is called the reference point and its deformation $\bm{u}(O)$ is set to zero for convenience.

In the case of the thermal problem, the use of periodic boundary conditions on temperature in the RVE also provides the
microscopic temperature distribution in an idealized medium formed as an infinite periodic arrangement of the RVE. In
this case, the corresponding equations are
\begin{equation} \label{pbcTemp}
  \theta(P^+_i) - \theta(P^-_i)
  = \nabla \bar{\theta}_i\cdot \bm{L}_i = \theta(M_i) - \theta(O)   \quad (i = 1,2, 3),
\end{equation}
where $\nabla \bar{\theta}$ is the temperature gradient of the macroscopic material point
and $\bm{L}_i,M_i,O$ are as before.

From the macroscopic view point, two cases of interest can be analyzed with the periodic RVE,
namely, adiabatic and  isothermal processes. In the case of an adiabatic process the average
temperature gradient $\nabla \bar{\theta}$ is equal to zero.  In this case, Eq.~\eqref{pbcTemp} is
reduced to
\begin{equation} \label{pbcTempAdia}
  \theta(P^+_i) - \theta(P^-_i) = 0   \quad (i = 1, 2, 3).
\end{equation}
In standard constitutive models at the macroscale, the adiabatic condition assumes that the plastic work contributes to
the macroscopic temperature increase locally and that its effect on the mechanical response is instantaneous. On the
contrary when two scales are considered, the heat generation at the micro-scale is not uniform due to the inhomogeneous
distributions of the stress, plastic strain and temperature fields caused by the heterogeneity of the microstructure.
In this case thermal conduction exists between grains at the microscopic level even under macroscopic adiabatic conditions
and the mechanical effect of the temperature at the macro-scale might not be instantaneous. Such a detailed description of
the distribution and evolution of the temperature field within the RVE is obtained in this
work by solving a thermo-mechanically coupled problem under the periodic boundary conditions given in Eqs.~\eqref{pbcDisp}
and ~\eqref{pbcTempAdia}. The additional complexity of the coupled problem leads to a better prediction
of the material behavior during thermo-mechanical processes, as compared to the conventional approaches.

The second case of interest is a macroscopic isothermal process. In this case, to ensure that the temperature of the
macroscopic material point does not change in time, the average temperature increase of the associated RVE should be kept
equal to zero at all times. This could be implemented simply by setting equal to zero the Taylor-Quinney factor $\chi$
defined in Eq.~\eqref{heatGen}.

\section{Numerical Methods}

  \subsection{Taylor model}
Under Taylor's homogenization model, the macroscopic deformation is
prescribed and it is assumed that its value is equal to the deformation gradient of each crystal.
Hence, there is no need to solve any boundary value problem at the microscale.  The problem is reduced
to solve, for each time increment, the system of non-linear equations defined by the constitutive
equations, summarized in section 2.2, and the Taylor approach given by Eq.~(\ref{eq:Taylor}).

To obtain the macroscopic stress and temperature as functions of the current deformation gradient,
the backward Euler method is used to integrate the equations in time. The time interval of interest
is divided into subintervals of size $\Delta t$. Let $\bm{\bar{F}}_{n+1}$ be the prescribed
macroscopic deformation gradient at time $t=t_{n+1}$ and $\bm{\sigma}^{(k)}$ the stress in
the $k-th$ grain of the polycrystal. The problem to solve consists in finding for each grain $k$ its
stress $\boldsymbol{\sigma}^{(k)}$, temperature, $\theta^{(k)}$ and internal variables
$\bm{\xi}^{(k)}$ that satisfy the constitutive equation~\eqref{eq-constitutive} at time
$t_{n+1}$, that is,
\begin{equation} \label{taylorStress}
\begin{split}
 \bm{\sigma}^{(k)}
  = \hat{\bm{\sigma}}
 \left( \bm{F}^{(k)}, \ \dot{\bm{F}}^{(k)}, \ \theta^{(k)}, \ \bm{\xi}^{(k)} \right)
 = \hat{\bm{\sigma}}
 \left( {\bar{\bm{F}}}_{n+1}, \ {\dot{\bm{\bar{F}}}}_{n+1}, \ \theta^{(k)}, \ \bm{\xi}^{(k)} \right).
\end{split}
\end{equation}
The numerical scheme for updating the local stress of each grain in Eq.~\eqref{taylorStress} is
discussed in detail in Section~\ref{subs-update-cauchy}.

The temperature at each grain, $\theta^{(k)}$ in Eq.~\eqref{taylorStress} is computed from
the local temperature at time $t_n$ and its increment from $t_n$ to $t_{n+1}$, denoted as $\Delta
\theta^{(k)}$, using the
thermal balance equation and neglecting heat conduction between grains 
\begin{equation} \label{taylortemp}
\Delta \theta^{(k)}=\dot{Q}_V^{(k)}(\boldsymbol{\sigma}^{(k)}, \bm{\xi}^{(k)}).
\end{equation}

A fixed-point iteration algorithm is used to solve, for each grain, the system of non-linear
equations~(\ref{taylorStress}) and (\ref{taylortemp}). The values of $\theta^{(k)}$ and
$\bm{\xi}^{(k)}$ are initialized at the beginning of the time step with the corresponding
values at the previous time and are updated iteratively. This iterative procedure is repeated until
the differences of the stress and temperature between two consecutive iterations are below a given
tolerances. 

Once the non-linear equations are solved, the macroscopic stress $\bar{\bm{\sigma}}$ and the
temperature $\bar{\theta}$ at the material point are computed as the average of the values in each
grain as
\begin{equation} \label{taylorHomo}
 \begin{split}
   \bar{\bm{\sigma}} &= \frac{1}{N_c} \sum^{N_c}_{k=1} \bm{\sigma}^{(k)}
    \\
   \bar{\theta} &= \frac{1}{N_c} \sum^{N_c}_{k=1} \theta^{(k)}
\end{split}
\end{equation}

\subsection{Computational homogenization approach}

\subsubsection{Iterative solution scheme for the coupled fields}

The governing equations of the coupled problem, namely Eqs.~\eqref{eq-balance-p} and
\eqref{eq-balance-e}, need to be satisfied simultaneously at any time.  For this combination of
nonlinear equations, a nonlinear incremental method is employed. In each increment, a two-level
iterative scheme is applied to solve the equations in a staggered way. First,
the deformation $\bm{\varphi}$ is found by solving the equilibrium equation~\eqref{eq-balance-p}
at a fixed temperature field. Then, the temperature $\theta$ is obtained as the solution
of the balance of energy equation at the newly updated deformation. The procedure
is repeated until both balance equations are solved below a given tolerance.

To solve for the deformation and temperature, a finite element discretization is employed
in space, and a backward Euler method is used to integrate the equations in time. For the
spatial discretization, we consider a partition of the body into a finite number of
hexahedra, defining a finite element space of shape functions $N_a:\mathcal{B}\to\mathbb{R}$,
where $a=1,\ldots,n_{node}$, and $n_{node}$ is the number of nodes in the mesh.
As in the case of the Taylor approach we partition the time interval of interest
into subintervals of size $\Delta t$, and indicate with the subscript $(\cdot)_n$
the value of any function at time $t_n= n\,\Delta t$. Using the
shape functions, the finite element approximation
of the deformation and temperature fields at time $t_n$,
denoted respectively as $\bm{\varphi}^h_n$
and $\theta^h_n$, can be expressed as
\begin{equation}
  \bm{\varphi}^h_n(\bm{X})
  = \sum_{a=1}^{n_{node}} N_a(\bm{X})\; \bm{\varphi}^a_n
  \ , \qquad
  {\theta}^h_n(\bm{X})
  = \sum_{a=1}^{n_{node}} N_a(\bm{X})\; {\theta}^a_n\ .
  \label{eq-fe}
\end{equation}
The quantities $\bm{\varphi}^a_n,\theta^a_n$ denote the \emph{nodal}
values of the deformation and temperature, respectively, at time $t_n$.

The Galerkin form of the equilibrium equation \eqref{eq-balance-p} is
obtained as follows. Given any admissible deformation variation $\delta\bm{\varphi}^h$,
the weak form of the balance of linear momentum can be written as
\begin{equation}
\begin{split}
  0
  =
  \int_{\mathcal{B}}
  \bm{\sigma} \cdot \nabla \delta\bm{\varphi}^h \,\mathrm{d} V
  -
  \int_{\partial \mathcal{B}}
  \bm{t} \cdot \delta\bm{\varphi}^h \,\mathrm{d} S
 -
 \int_{\mathcal{B}} \bm{b} \cdot \delta\bm{\varphi}^h \,\mathrm{d} V ,
\end{split}
\label{eq-fem-p}
\end{equation}
where $\bm{\sigma}$ is the value of the Cauchy stress
obtained at the solution of the plasticity model of
Sections~\ref{subs-cp} and \ref{subs-homo}, evaluated  at time $t_{n+1}$
as detailed in Section~\ref{subs-update-cauchy}.

Similarly, for a test temperature field $\delta{\theta}^h$,
the backward Euler discretization of the weak form
of Eq.~\eqref{eq-balance-e} reads
\begin{equation}
  \begin{split}
    \int_{\mathcal{B}}
    \left[
    \rho c \frac{\theta^h_{n+1} - \theta^h_n}{\Delta t}
    \delta\theta^h
    +
    \nabla\theta^h_{n+1} \cdot \bm{k}  \nabla\delta\theta^h
    -
    \dot{Q}_V\;  \delta\theta^h
    \right]
    \,\mathrm{d} V
    = 0 .
  \end{split}
\label{eq-fem-e}
\end{equation}

As indicated previously, we have implemented the solution of Eqs.~\eqref{eq-fem-p}-\eqref{eq-fem-e}
in a staggered fashion: first, Eq.~\eqref{eq-fem-p}
is solved for $\bm{\varphi}^h_{n+1}$, for a fixed temperature; then, Eq.~\eqref{eq-fem-e}
is used to find the updated temperature $\theta^h_{n+1}$, for the previously
obtained deformation. The two steps are repeated until both discrete balance
equations are verified up to a given tolerance.

\subsubsection{Update of Cauchy stress}
\label{subs-update-cauchy}
The update of the stress tensor is the key part of the iterative process required
to solve the nonlinear system of coupled equations. During each iteration,
the thermal deformation gradient can be calculated directly from Eq.~\eqref{evolveFTheta} as
\begin{equation} \label{FTheta}
  \bm{F}^\theta_{n+1} = \exp \left( \Delta \theta \bm{\beta} \right) \bm{F}^\theta_n
  \approx \left( \bm{I} + \Delta \theta \bm{\beta} \right) \bm{F}^\theta_n ,
\end{equation}
with $\Delta \theta = \theta_{n+1}-\theta_n$.
Next, using Eq.~\eqref{defGrdDcmp}, the plastic deformation gradient can be evaluated
with the expression
\begin{equation} \label{FPlas}
  \bm{F}^p_{n+1}
  =
  \exp \left( \Delta t \sum_{\alpha} {\dot{\gamma}^\alpha   \bm{S}^{\alpha}_0} \right)
  \bm{F}^p_n
  \approx \left( \bm{I} + \Delta t \sum_{\alpha} {\dot{\gamma}^\alpha \bm{S}^\alpha_0} \right)
  \bm{F}^p_n .
\end{equation}
Replacing Eqs.~\eqref{FTheta} and~\eqref{FPlas} in Eq.~\eqref{defGrdDcmp},
the elastic deformation gradient can be also obtained as
\begin{equation}
  \bm{F}^e_{n+1} =
  \bm{F}_{n+1} {\bm{F}^\theta_{n+1}}^{-1} {\bm{F}_n^p}^{-1}
  \left( \bm{I} - \Delta t \sum_{\alpha} {\dot{\gamma}^\alpha \bm{S}^\alpha_0} \right) .
  \end{equation}
This expression can be rewritten as
\begin{equation} \label{resdFElas}
   \bm{F}^e_{n+1} - \bm{F}_{n+1}
   {\bm{F}_{n+1}^\theta}^{-1} {\bm{F}_n^p}^{-1}
   \left( \bm{I} - \Delta t \sum_{\alpha} {\dot{\gamma}^\alpha \bm{S}^\alpha_0} \right) = 0 .
\end{equation}
Using an Euler backward scheme, the increment of the accumulated shear strain is given by
\begin{equation} \label{resdGammaInc}
   \Delta \gamma^\alpha = \dot{\gamma}^\alpha \Delta t .
\end{equation}
Combining Eqs.~\eqref{resdFElas} and~\eqref{resdGammaInc},
a set of nonlinear equations is formed, and the elastic deformation gradient $\bm{F}^e_{n+1}$ and the increment of the accumulated
shear strain $\Delta \gamma^\alpha$ can be solved simultaneously, at the level of the quadrature
point, with Newton-Raphson's method. Once the elastic deformation gradient of the current
increment is obtained, the Cauchy stress is updated with Eqs.~\eqref{GreenStrain}-\eqref{CauchyStress}.

\section{Numerical Examples}

The computational framework proposed is applied to study the thermo-mechanical
response of pure BCC tantalum under uniaxial traction. The results obtained with this numerical framework
will be compared at the end of this section with the response obtained using the Taylor
homogenization approach.

\begin{figure}[th]\centering
    \includegraphics[width=0.45\textwidth]{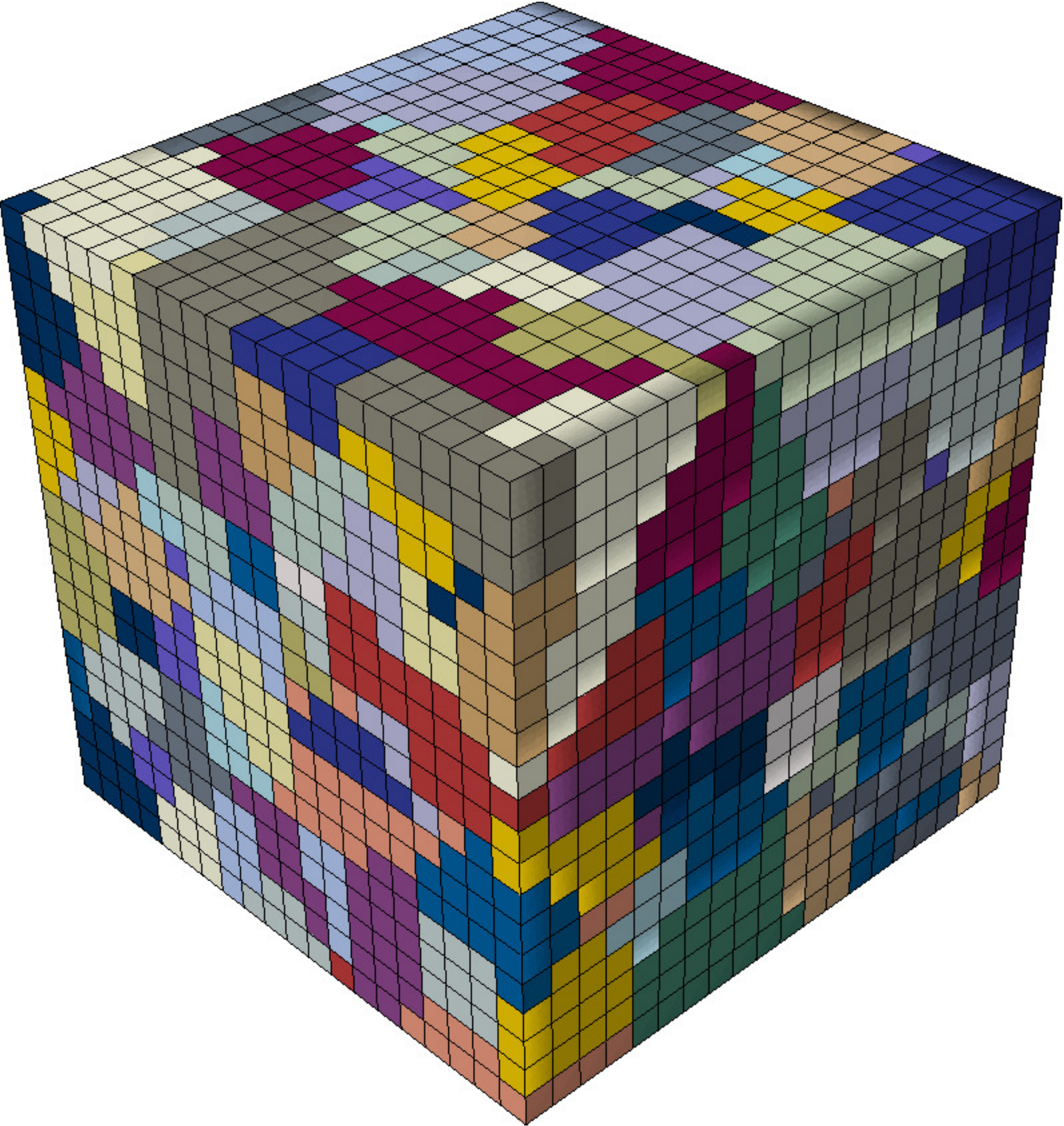} \\
    \caption{Representative volume element of BCC Ta containing
      400 randomly-oriented crystals discretized with 8000 cubic finite elements.}
    \label{RVE}
  \end{figure}

A RVE with 400 randomly oriented grains is defined and discretized in hexahedral finite elements, as
illustrated in Fig.~\ref{RVE}. To  determine the size of the models a mesh sensitivity study was
done to. In particular, simulations considering the macroscopic adiabatic condition were conducted
using meshes containing $20^3, 25^3$ and $30^3$ elements to check the effects of mesh. It was
observed that the difference of both the macroscopic temperature-strain curves and the stress-strain
curves between the model with $20^3$ mesh and $30^3$ mesh were below 1\%. In addition, microscopic
fields were also very similar. These results suggested that the mesh containing $20^3$ elements is
accurate enough, and this size is then used for the following part of the paper
(Fig.~\ref{RVE}). With respect to the number of crystals,  it has been observed that 400 is a
number sufficiently large to provide statistically representative results and isotropic response.

The elastic moduli (in GPa) of this metallic material is approximated by a linear function of the
temperature $\theta$ as follows \cite{Simmons1971}:
  \begin{equation}
    \begin{split}
       & c_{11} =  268.2 - 0.024\,\theta,  \\
       & c_{12} =  159.6 - 0.011\,\theta,  \\
       & c_{44} = \ 87.1 - 0.015\,\theta.  \\
    \end{split}
  \end{equation}
The density, the specific heat, the thermal conductivity and the thermal expansion are taken to be $16.69\times 10^3$ Kg/m$^3$,
$1.506\times 10^8$ J/(Kg $\cdot$ K), 57.5 W/(m $\cdot$ K) and $6.3\times 10^{-6}$ K$^{-1}$, respectively.

Regarding the plastic response, in BCC metals three sets of slip systems are normally responsible of the plastic deformation,
$\langle 1 \, 1 \, 1 \rangle \left\{1 \, 1 \, 0\right\}$, $\langle 1 \, 1 \, 1 \rangle \left\{1 \, 1 \, 2\right\}$ and
$\langle 1 \, 1 \, 1 \rangle \left\{1 \, 2 \, 3\right\}$. In this work, only the first two slip sets are taken into
consideration as proposed in \cite{KOTHARI199851} and the corresponding slip systems are listed in Table~\ref{slipSystems}.
\begin{table}[ht]
  \small
  \centering
  \caption{Slip Systems for BCC Ta}
  \label{slipSystems}
  \begin{tabular}{ |c|c|c|c|c|c| } \hline
    $\alpha$   &    $\bm{s}_0$       &    $\bm{m}_0$      &   $\alpha$     &   $\bm{s}_0$     &   $\bm{m}_0$      \\  \hline
    1          &    1 \, 1 \, $\bar{1}$    &    0 \, 1 \,         1   &   13           &   $\bar{1}$ \, 1 \, 1  &   2 \, 1 \,         1   \\  \hline
    2          &    1 \, $\bar{1}$ \, 1    &    0 \, 1 \,         1   &   14           &   1 \, 1 \,         1  &   $\bar{2}$ \, 1 \, 1   \\  \hline
    3          &    $\bar{1}$ \, 1 \, 1    &    0 \, 1 \, $\bar{1}$   &   15           &   1 \, 1 \, $\bar{1}$  &   2 \, $\bar{1}$ \, 1   \\  \hline
    4          &    1         \, 1 \, 1    &    0 \, 1 \, $\bar{1}$   &   16           &   1 \, $\bar{1}$ \, 1  &   2 \, 1 \, $\bar{1}$   \\  \hline
    5          &    1 \, 1 \, $\bar{1}$    &    1 \, 0 \,         1   &   17           &   1 \, $\bar{1}$ \, 1  &   1 \, 2 \,         1   \\  \hline
    6          &    $\bar{1}$ \, 1 \, 1    &    1 \, 0 \,         1   &   18           &   1 \, 1 \, $\bar{1}$  &   $\bar{1}$ \, 2 \, 1   \\  \hline
    7          &    1 \, $\bar{1}$ \, 1    &    1 \, 0 \, $\bar{1}$   &   19           &   1 \, 1 \,         1  &   1 \, $\bar{2}$ \, 1   \\  \hline
    8          &    1 \,        1  \, 1    &    1 \, 0 \, $\bar{1}$   &   20           &   $\bar{1}$ \, 1 \, 1  &   1 \, 2 \, $\bar{1}$   \\  \hline
    9          &    $\bar{1}$ \, 1 \, 1    &    1 \, 1 \,         0   &   21           &   1 \, 1 \, $\bar{1}$  &   1 \, 1 \,         2   \\  \hline
    10         &    1 \, $\bar{1}$ \, 1    &    1 \, 1 \,         0   &   22           &   1 \, $\bar{1}$ \, 1  &   $\bar{1}$ \, 1 \, 2   \\  \hline
    11         &    1 \, 1 \, $\bar{1}$    &    1 \, $\bar{1}$ \, 0   &   23           &   $\bar{1}$ \, 1 \, 1  &   1 \, $\bar{1}$ \, 2   \\  \hline
    12         &    1 \, 1  \,        1    &    1 \, $\bar{1}$ \, 0   &   24           &   1 \, 1 \,         1  &   1 \, 1 \, $\bar{2}$   \\  \hline
  \end{tabular}
\end{table}
The experimental behavior of Tantalum was taken from \cite{NEMATNASSER1997907} where an enhanced compression recovery split Hopkinson bar technique was used to obtain the plastic flow in wide range of strains, strain rates, and temperatures during uniaxial compression. In particular, the model parameters defining both flow and hardening model were adjusted to fit the stress-strain response of the polycrystalline RVE with the experimental result for the case of uniaxial traction under isothermal condition in \cite{NEMATNASSER1997907}.
This fitting procedure used in the study started using the set of parameters for the flow and hardening model in \cite{KOTHARI199851} as an initial guess. It must be noted that the resulting parameters are different to the ones obtained in \cite{KOTHARI199851}  because their polycrystalline framework was based on Taylor?s model, which gives an upper bound for the stress predictions, while the present approach is based on computational homogenization. The resulting  set of parameters obtained are listed in Table~\ref{modelParameters}.
    \begin{table}[ht]
      \small
      \centering
      \caption{Material Properties of BCC Ta}
      \label{modelParameters}
        \begin{tabular}{ |l|l| } \hline
        Model Parameter       &   Value                              \\
                   \hline
        ${\Delta}F^{\alpha}$  &   $2.77\cdot10^{-19}$ J              \\
        $\dot{\gamma}_0$      &   $1.732\cdot10^{7}$ s$^{-1}$        \\
        $h_0$                 &   181 MPa                            \\
        $s_s$                 &   121 MPa                            \\
        $r_1$                 &   1.10                               \\
        $p$                   &   0.28                               \\
        $q$                   &   1.40                               \\
        $s_{at, 0}$           &   50.0 MPa                           \\
        $s_{t, 0}$            &   445  MPa                           \\
                   \hline
      \end{tabular}
    \end{table}

\begin{figure}[htbp]\centering
  \includegraphics[width=0.45\textwidth]{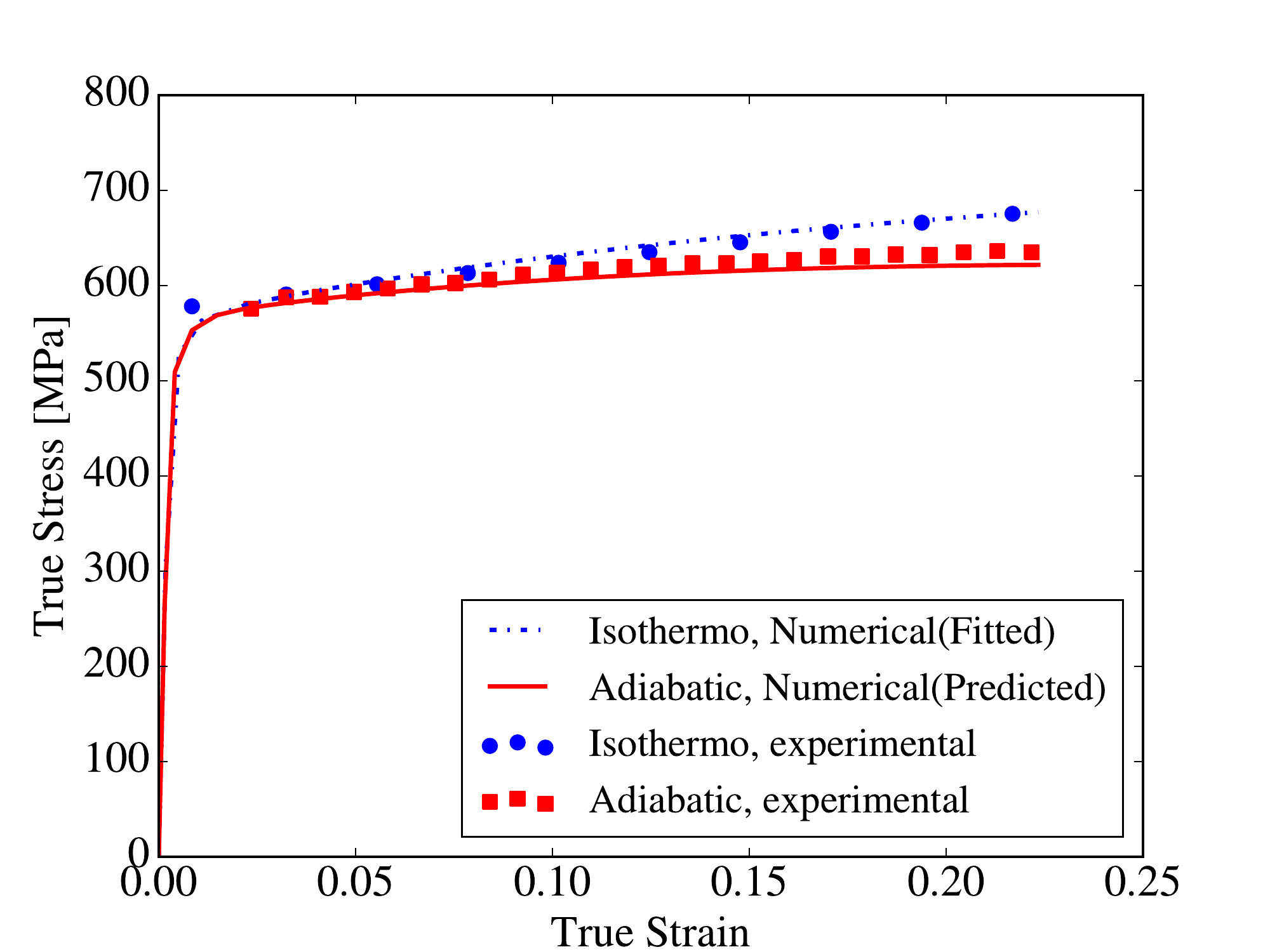} \\
  \caption{Comparison of experimental and predicted stress-strain responses of BCC Ta during uniaxial traction.}
  \label{Fitting}
\end{figure}

To assess the validity of the fitting procedure, the parameters listed in Table~\ref{modelParameters} (fitted using isothermal experiments) are used in the numerical framework to predict the stress-strain response of the RVE under an adiabatic condition. The numerical results and corresponding experimental results
obtained from \cite{NEMATNASSER1997907} are plotted in Fig.\,\ref{Fitting}. It is shown that the numerical
predictions provided by the current work are in good agreement with the experimental results, which validates
the reliability of this proposed framework.

\subsection{Effects of initial orientation}

\begin{figure*}[ht]\centering
    \subfigure[]{
      \label{S33Ori}
      \includegraphics[width=0.45\textwidth]{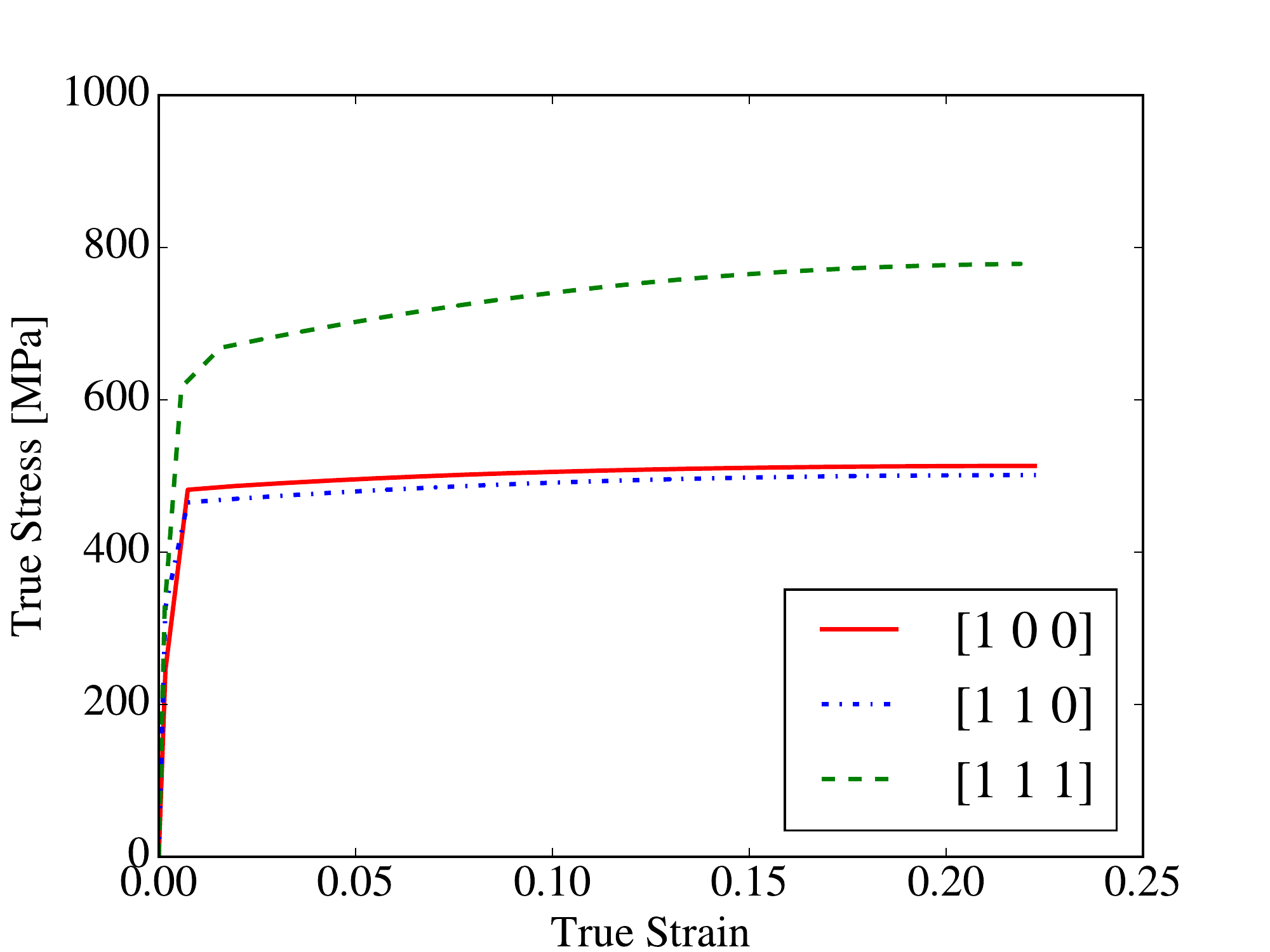}
    }
    \hfill
    \subfigure[]{
      \label{TempOri}
      \includegraphics[width=0.45\textwidth]{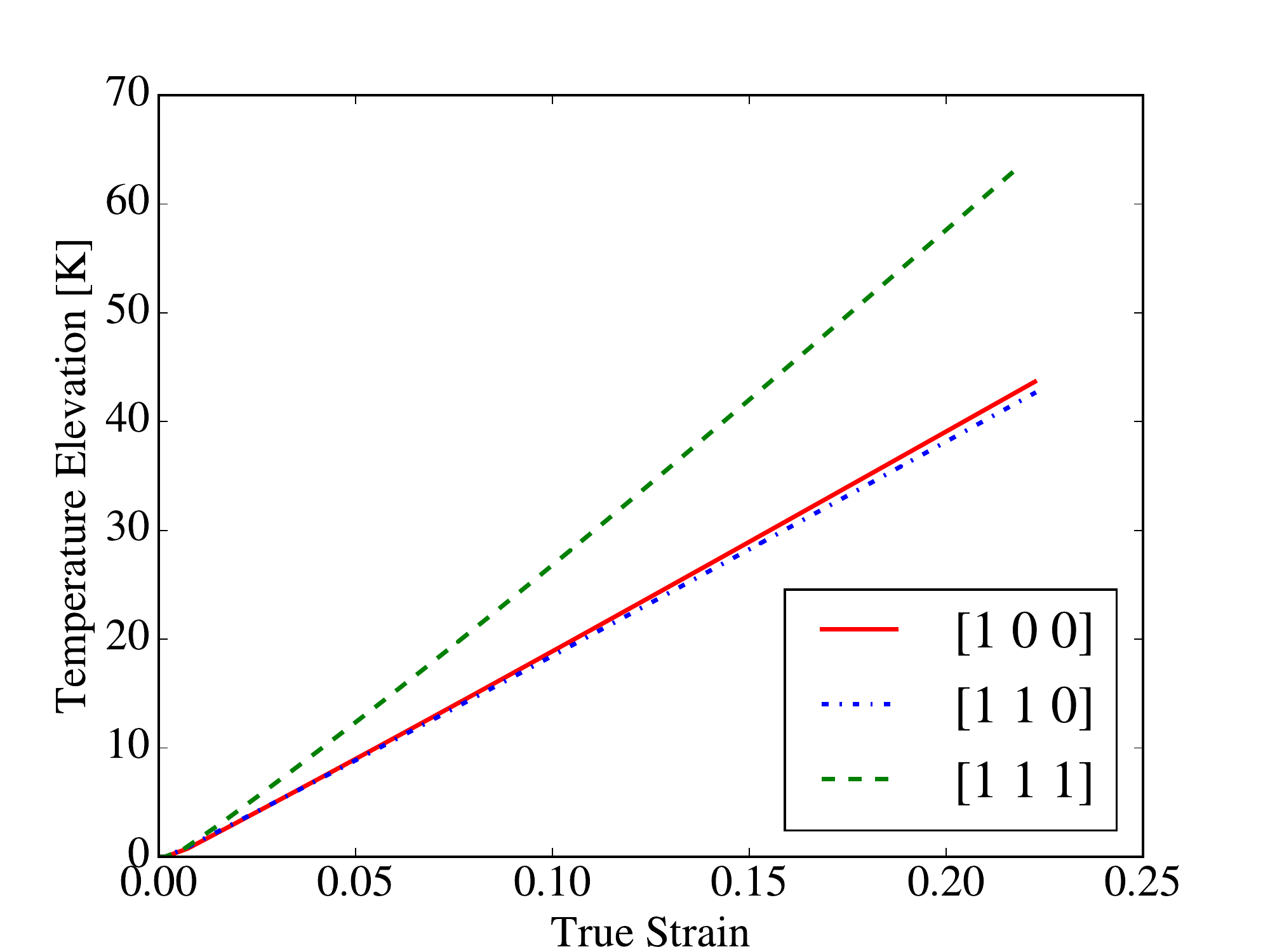}
    }
    \caption{(a). Uniaxial stress-strain and (b). Temperature
      change-strain curves of single crystalline BCC Ta during uniaxial traction at various crystallographic directions}
    \label{Orientation}
  \end{figure*}

To study the effect of the initial orientation on the responses of single crystal Ta, a single
crystal RVE is created and discretized into $20\times20\times20$ regular finite elements, as in the
first simulation. The solid is strained at three
different crystalline directions [1 0 0], [1 1 0] and [1 1 1]. The uniaxial stress-strain curves and
the temperature changes at these directions during traction processes under adiabatic condition are plotted in
Fig.~\ref{Orientation}.

It is shown that the response of the single crystal is closely dependent on the traction direction, being the strength
obtained when loading in direction [1 1 1] the highest of all the three cases. In the model, the same value of $s_{at}$
and $s_t$ are chosen for all the slip systems so the result obtained is simply due to the Schmidt factor of the systems.
In particular, when loading in a [1 1 1] direction the Schmidt factor of the best oriented planes is lower than in the
other two loading directions and therefore, the material plastifies at a larger applied stress.

The implications of the loading direction of the single crystal in the thermal response can be observed in Fig.~\ref{Orientation}.
The temperature rises faster when loading in the [111] direction and this is due to the higher strength in this direction.
Heat generation is proportional to the plastic dissipation which {rate} depends on the sum of the product of resolved stress
and shear rates on all the systems (Eq.~\eqref{taylorquinney_CP}). Because the accumulated plastic shear is the same for
all the three loading directions, the temperature increases faster when pulling in the [111] direction due to its larger strength.

\subsection{Effects of initial temperature}

\begin{figure*}[ht]\centering
    \subfigure[]{
      \label{S33Temp}
      \includegraphics[width=0.45\textwidth]{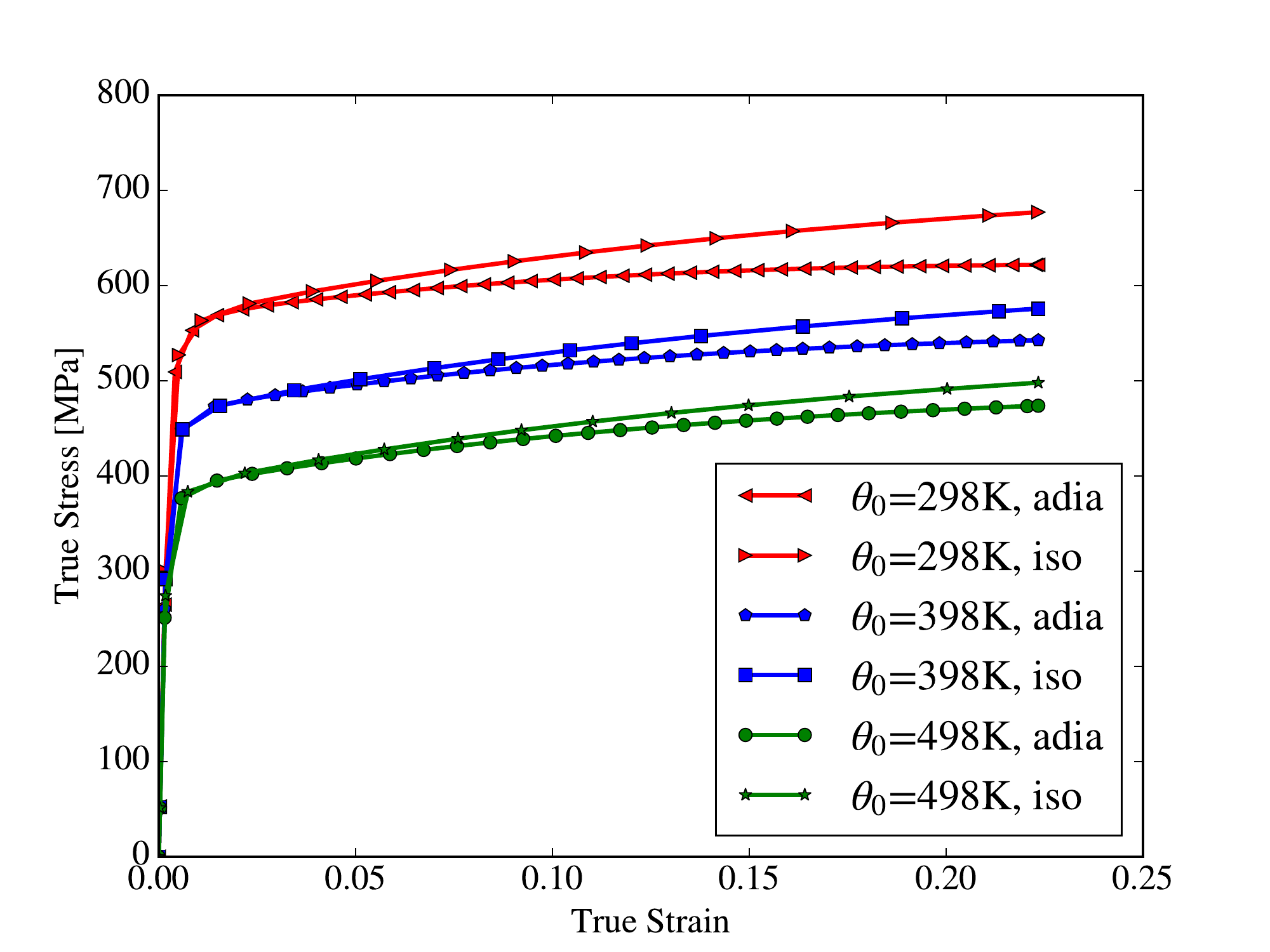}
    }
    \hfill
    \subfigure[]{
      \label{TempTemp}
      \includegraphics[width=0.45\textwidth]{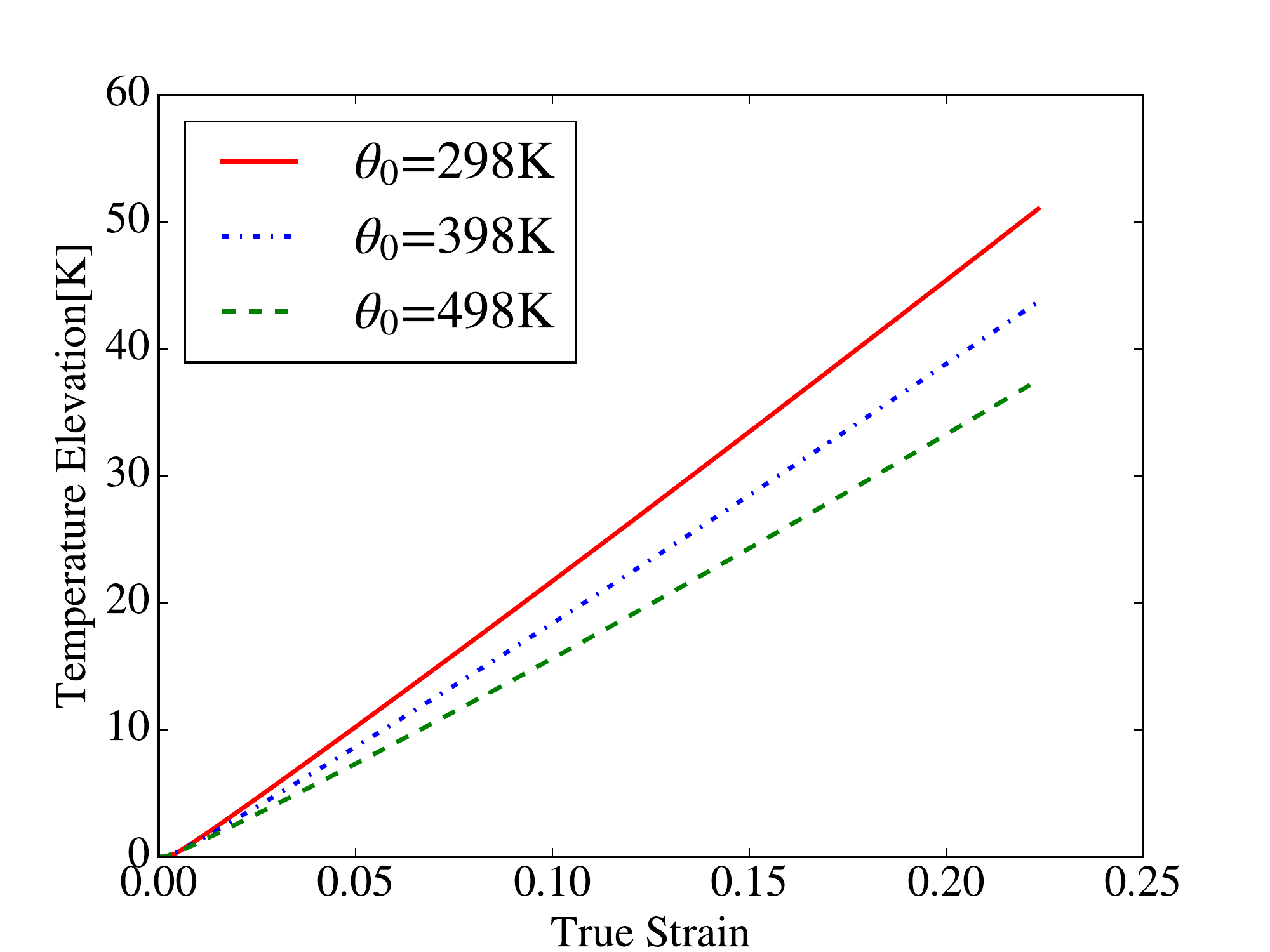}
    }
    \caption{(a). Stress- (b). Temperature change-strain curves of
      BCC Ta during uniaxial traction at various initial temperatures}
    \label{Temp}
  \end{figure*}

To study the effect of the initial temperature on the thermo-mechanical behavior of polycrystalline Ta, the RVE
is strained at three different initial temperatures: 298 K, 398 K, and 498 K with an engineering strain
rate of 5000 $s^{-1}$. Two sets of simulations are performed, under isothermal conditions and under adiabatic conditions
in order to quantify the effect of considering heat generation during the test. The numerical results of the mechanical
response of both type of tests are plotted in Fig.~\ref{Temp}(a). It can be observed that the effect of the initial
temperature is a decrease in the material strength. In the isothermal case, all the three curves are proportional and
hardening is not affected by the initial temperature. Under adiabatic conditions, the material plastifies at the same
stress than as in the corresponding isothermal case but the stress-strain curve deviates with the strain showing less
hardening. Moreover, in contrast with the isothermal case, the effective hardening rate increases when decreasing the initial temperature. The reason of this behavior can be found in the heat generation due to plastic dissipation. As it can be
observed in  Fig.~\ref{Temp}(b) the temperature rise is higher for the lowest temperature due to the higher value of
the strength (as it happened when analyzing the effect of loading direction). This temperature increase produces a softening of the material that is more evident for the lower initial temperature where the differences between isothermal and adiabatic cases are maximal. The resulting behavior is in agreement with the experimental observations showing a reduction of the strength for BCC Ta as \uline{a} function
of the initial temperature.

\subsection{Effects of strain rate}
\begin{figure*}[htbp]\centering
    \subfigure[]{
      \label{S33ERateAdia}
      \includegraphics[width=0.45\textwidth]{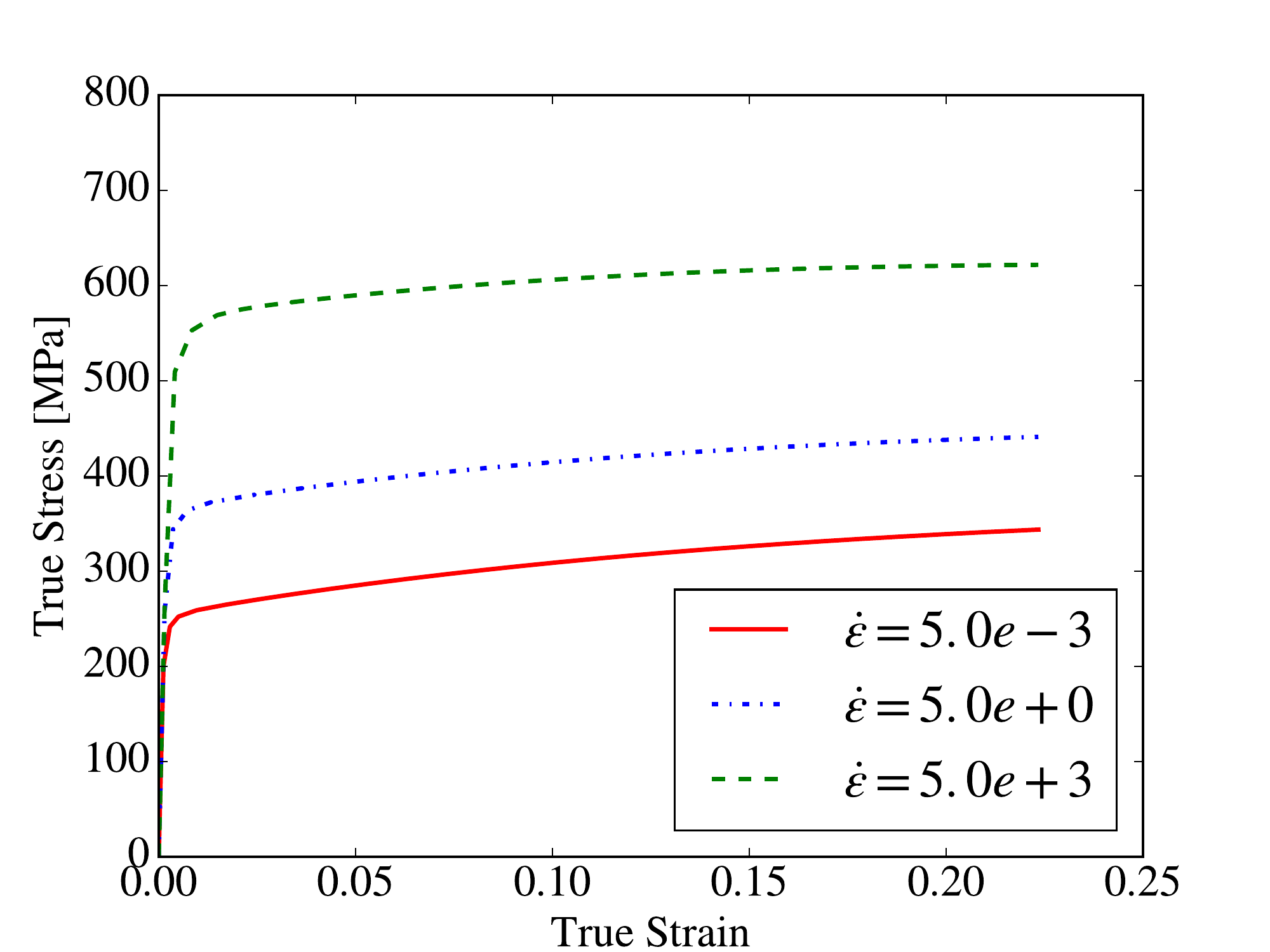}
    }
    \hfill
    \subfigure[]{
      \label{TempERateAdia}
      \includegraphics[width=0.45\textwidth]{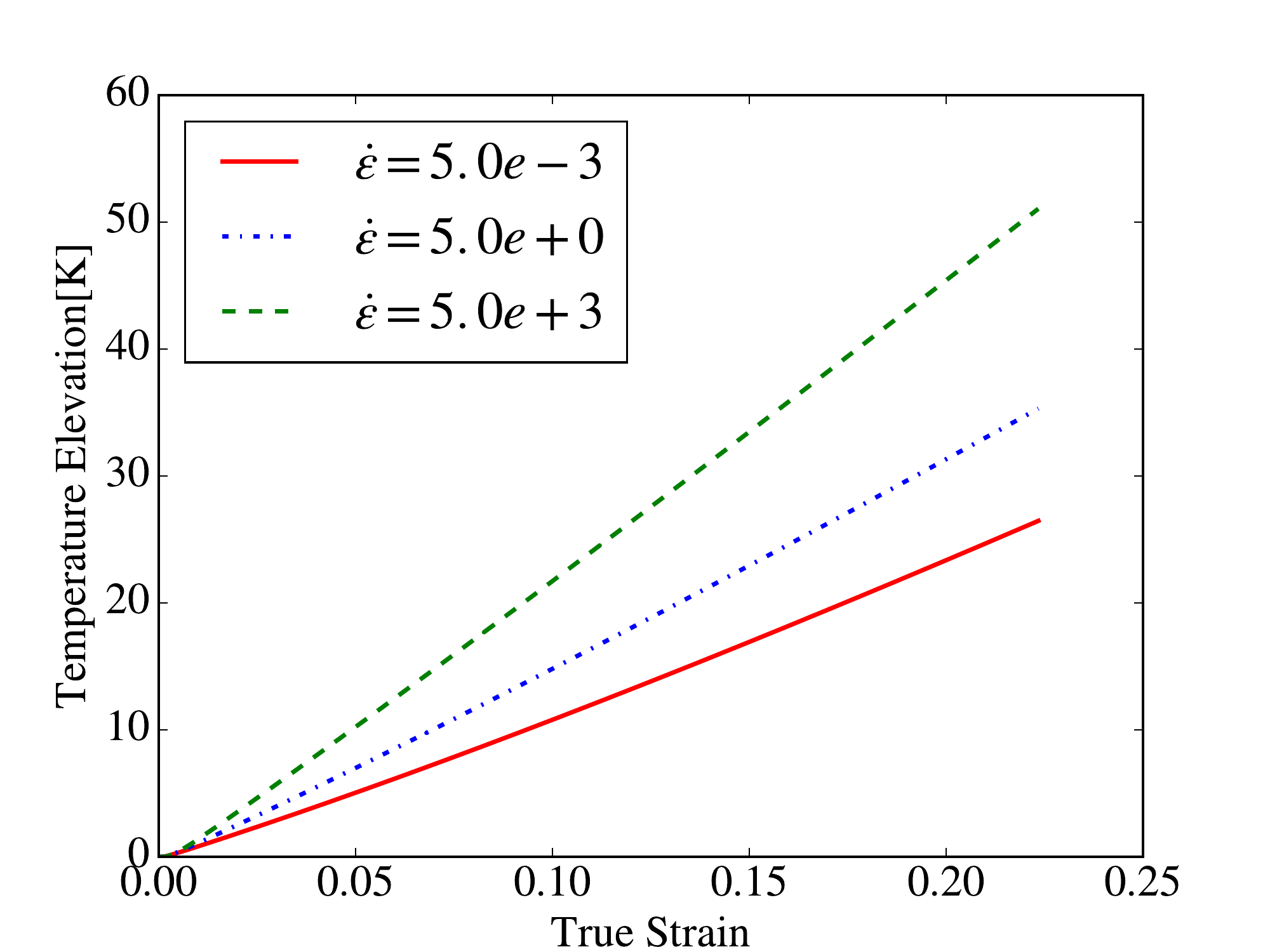}
    }
    \caption{(a). Stress- (b). Temperature change-strain curves of BCC Ta during uniaxial traction
      tests at various engineering strain rates}
    \label{ERateAdia}
  \end{figure*}

To study the effect of the strain rate on the material behavior of polycrystal Ta, the RVE is
strained at three different engineering strain rates: $5\cdot10^{-3}$, $5$, and $5\cdot10^3$
s$^{-1}$, all of them at 298K and under adiabatic conditions.
The stress-strain response of the polycrystal for the three strain rates considered are plotted in Fig.~\ref{ERateAdia}(a).
In this figure the effect of the strain rate can be clearly observed, the strength of the polycrystal increases when
increasing the strain rate applied. These results are in quantitative agreement with the experimental
results in \cite{KOTHARI199851}. The hardening effect is due to the elasto-visco-plastic nature of the CP selected and
 shows  that increasing the strain rate is equivalent to   reducing  the temperature. The large differences found in the
initial yield   indicate a high strain rate sensitivity of the model for the energy parameters chosen even at room
temperature. The effect of the strain rate in the thermal response of the polycrystal can be observed in Fig.~\ref{ERateAdia}(b)
where the temperature increase is represented as a function of the applied strain for the three strain rates considered.
The larger strain rates result in higher values of the effective strength and as a consequence the plastic dissipation
increases and the temperature increases faster. It is
interesting to note that the current framework is able to reproduce the analogous effect of decreasing the applied strain
rate or increasing the initial temperature in both the mechanical response and in the temperature evolution.

\subsection{Effects of considering heat conduction at the microscale}

Conventional approaches of thermo-mechanical coupled crystal plasticity often neglect the heat
transfer.  To study the effect of heat transfer on the responses of BCC Ta, comparative simulations
were carried out to study the responses of the RVE during uniaxial traction tests , under
macroscopic adiabatic conditions.  Two different cases are considered, one where heat conduction at
the microscale is considered and a second one where heat transfer is disabled also at the
microscopic level. The maximum, minimum, and average value of the microscopic
temperature field obtained for the three strain rates considered are given in
Tab. \ref{tempDistCmp}.
\begin{table}[ht]
  \small
  \centering

  \begin{tabular}{ |c|c|c|c|c|c|c| } \hline
    \multirow{2}{*}{}   &  \multicolumn{2}{|c|}{$\dot{\varepsilon}=5000$ s$^{-1}$}   &   \multicolumn{2}{|c|}{$\dot{\varepsilon}=5$ s$^{-1}$} &  \multicolumn{2}{|c|}{$\dot{\varepsilon}=0.005$ s$^{-1}$}   \\
    \cline{2-7}         &   No diff.  &   Full model  &   No diff.   &   Full model    &   No diff.  &   Full model  \\   \hline
    $T_{min}$ (K)       &   304.2    &   316.3    &   302.3    &    333.2    &   300.9     &    324.5      \\   \hline
    $T_{max}$ (K)       &   422.8    &   401.2    &   385.1    &    333.5    &     364.0   &    324.5      \\   \hline
    $T_{ave}$ (K)       &   349.1    &   349.1    &   333.2     &    333.3    &      324.4   &    324.5      \\   \hline
  \end{tabular}  \caption{Maximum, minimum, and average local temperatures considering heat diffusion (full model) or a microscopic adiabatic condition (no diffusion)}  \label{tempDistCmp}
\end{table}
When heat diffusion is considered, it can be observed that for the two smallest strain
rates, namely 5s$^{-1}$ and 5 \ $10^{-3}$s$^{-1}$, the temperature distribution on the microstrucure is
homogeneous. In contrast, when microscopic heat conduction is not allowed (microscopic adiabatic
condition), temperature distributions are heterogeneous and the differences between the maximum and
minimum local temperatures are 80K and 63K, respectively, for the two strain rates. In contrast,
for the highest strain rate (5\ $10^{3}$s$^{-1}$) the full model with diffusion predicts an
heterogeneous temperature distribution. This is an important result of the model because it shows
that a characteristic time scale is introduced. Then, when strain rates are low, temperature becomes
homogeneous, while for fast deformations the material behaves almost as microscopic-adiabatic.

\begin{figure*}[ht]\centering
    \subfigure[]{
      \label{tempER5000WtCond}
      \includegraphics[width=0.45\textwidth]{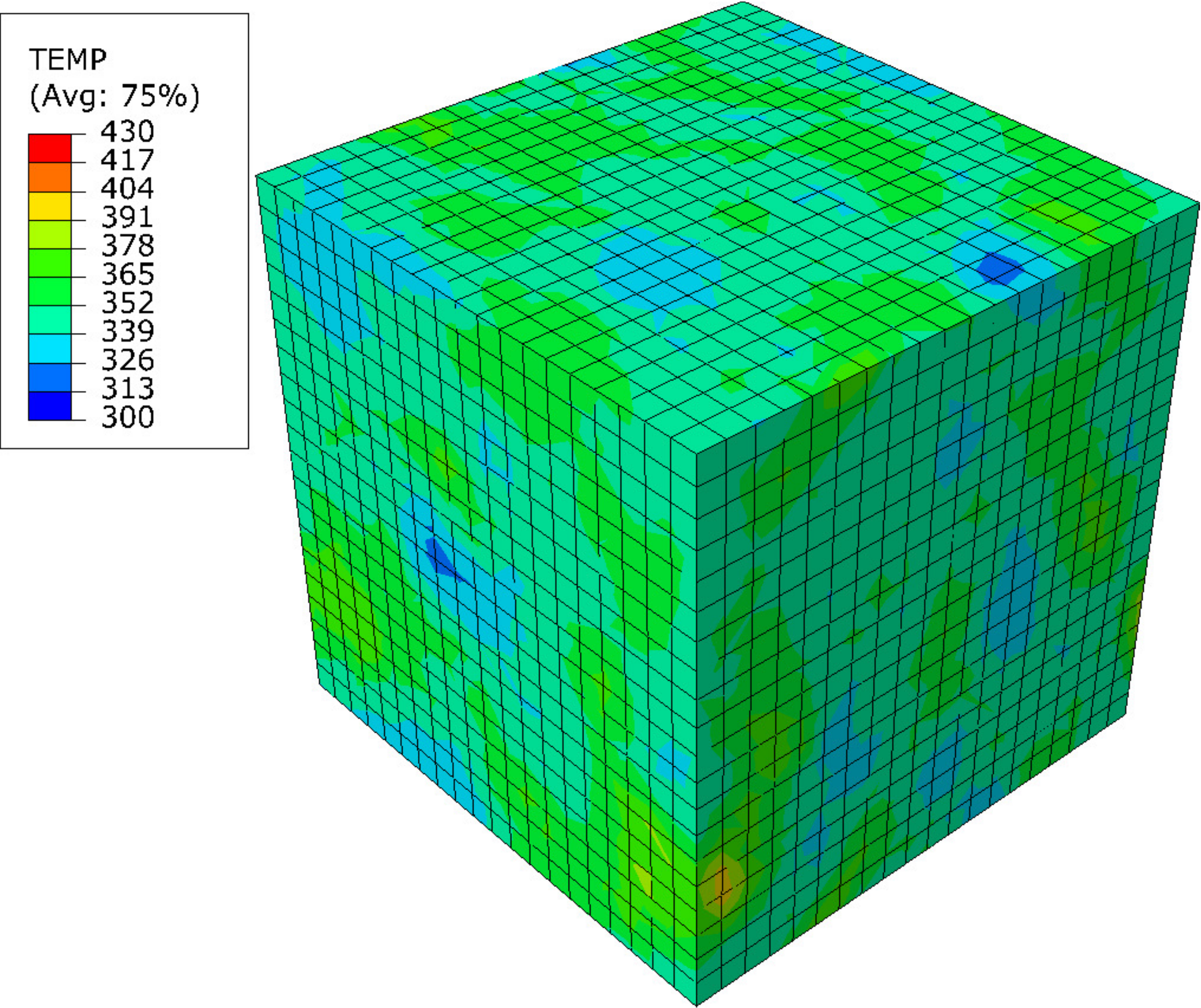}
    }
    \hfill
    \subfigure[]{
      \label{tempER5000WoCond}
      \includegraphics[width=0.45\textwidth]{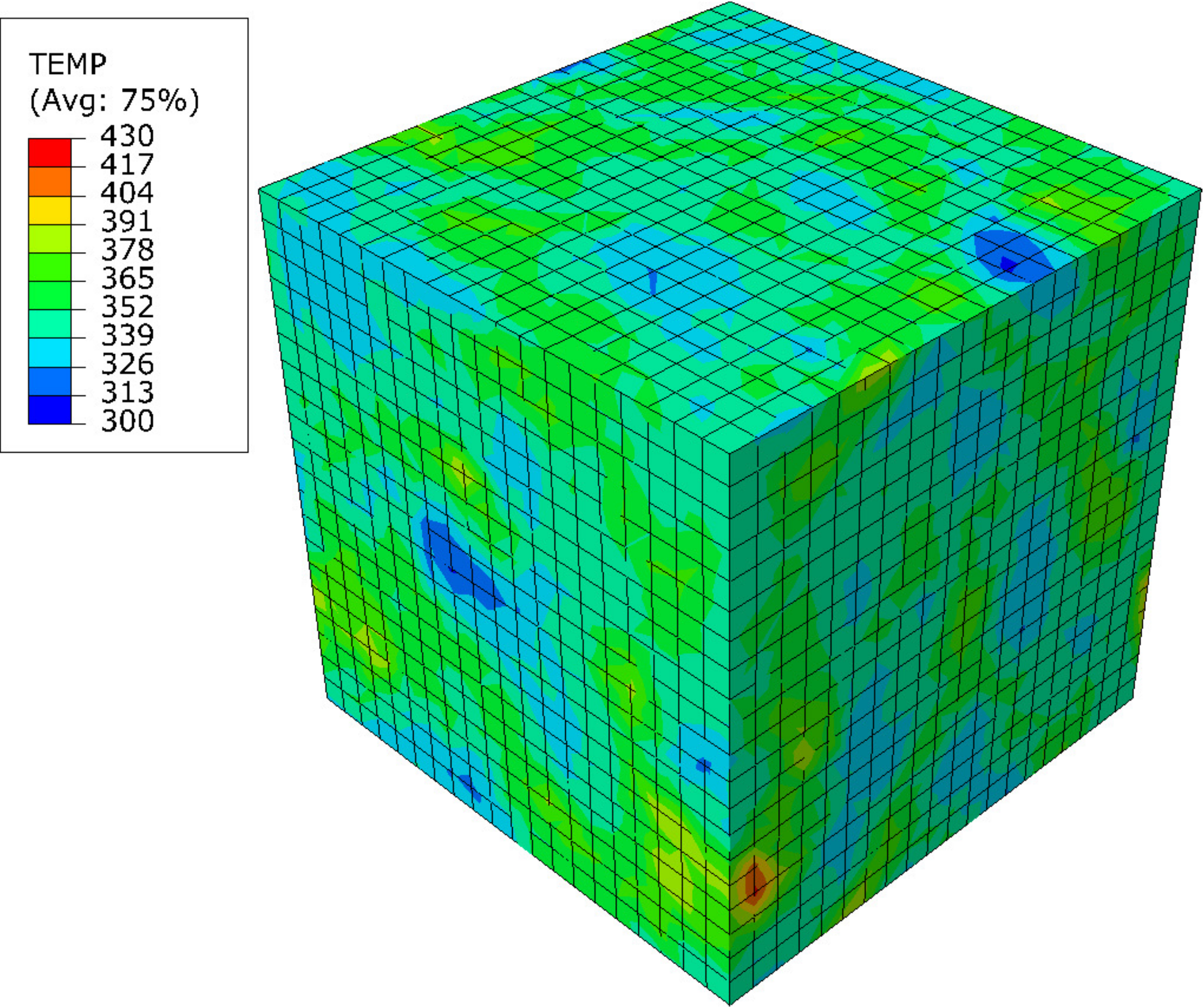}
    }
    \caption{Temperature distributions of BCC Ta during
      uniaxial traction tests at $\dot{\varepsilon}=5.0\cdot10^{3}$ s$^{-1}$
      under cases (a). with heat transfer and
     (b). without heat conduction }
    \label{effHeatCond}
  \end{figure*}

To further analyze this behavior, the temperature field distribution at the end of the tensile
simulation for the case at a nominal strain rate of 5000~s$^{-1}$ is plotted in
Fig.~\ref{effHeatCond}(a) for the case in which thermal conduction is considered and (b) for the
microscopic-adiabatic case. Both figures show an heterogeneous distribution of the temperature due
to the non-homogeneous distribution of the plastic deformation within the RVE. In both cases, the
temperature hot-spots are the points in which plastic dissipation is maximum and correspond to the
regions in which plastic strain is localized, near grain boundary or triple points. The comparison
between the two figures suggests that the heat transfer plays a smoothing role, as
expected. Allowing heat transfer at the microscale diffuses the temperature near the hottest points
of the microstructure and the temperature gradients in Fig.~\ref{tempER5000WtCond} are reduced in
comparison to the case without transfer, as shown in Fig.~\ref{tempER5000WoCond}. This
redistribution is very incipient here (time is 5$10^{-5}$s) and temperature heterogeneities are
still significant, but the heterogeneities disappear for larger times, as it happens at the end of the tests
at 5s$^{-1}$ and 5$10^{-3}$s$^{-1}$ (Tab \ref{tempDistCmp}).

  \subsection{Comparison with Taylor homogenization model}

To illustrate the difference between the  results obtained using the full-field computational homogenization
approach proposed in this work and the Taylor model, the cases studied in the previous sections are now
simulated using this last approximation. In particular, the model was used to simulate uniaxial
tensile tests performed at nominal strain rates of 0.005, 5.0 and 5000~s$^{-1}$, under both
macroscopic adiabatic and isothermal conditions. The initial temperature was chosen to be 298~K in
all cases.

\begin{figure*}[ht]\centering
  \hfill
  \subfigure[]{
    \label{S33ER5000IsoCmp}
    \includegraphics[width=0.45\textwidth]{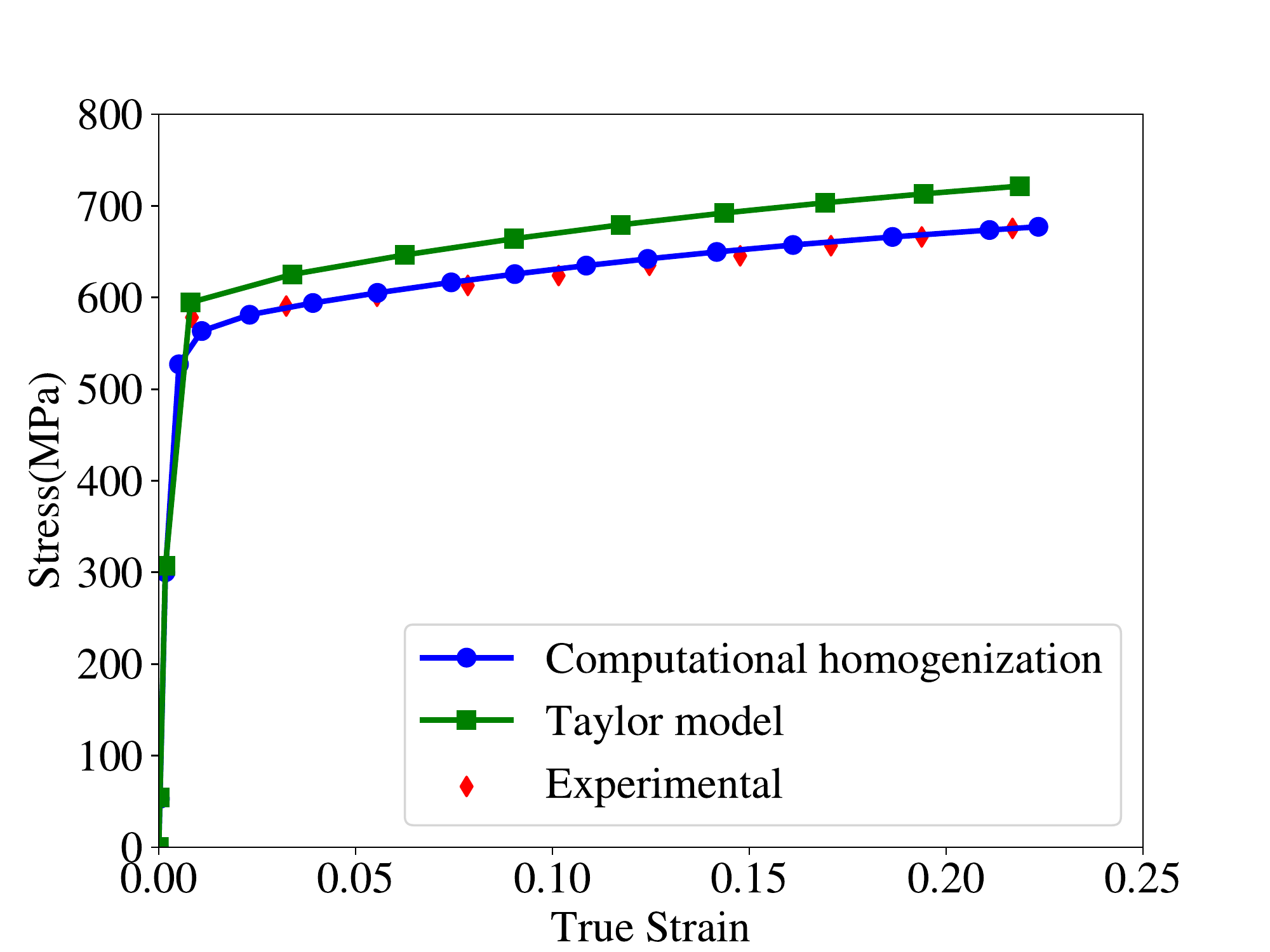}
  }
  \hfill
  \subfigure[]{
    \label{S33ER5IsoCmp}
    \includegraphics[width=0.45\textwidth]{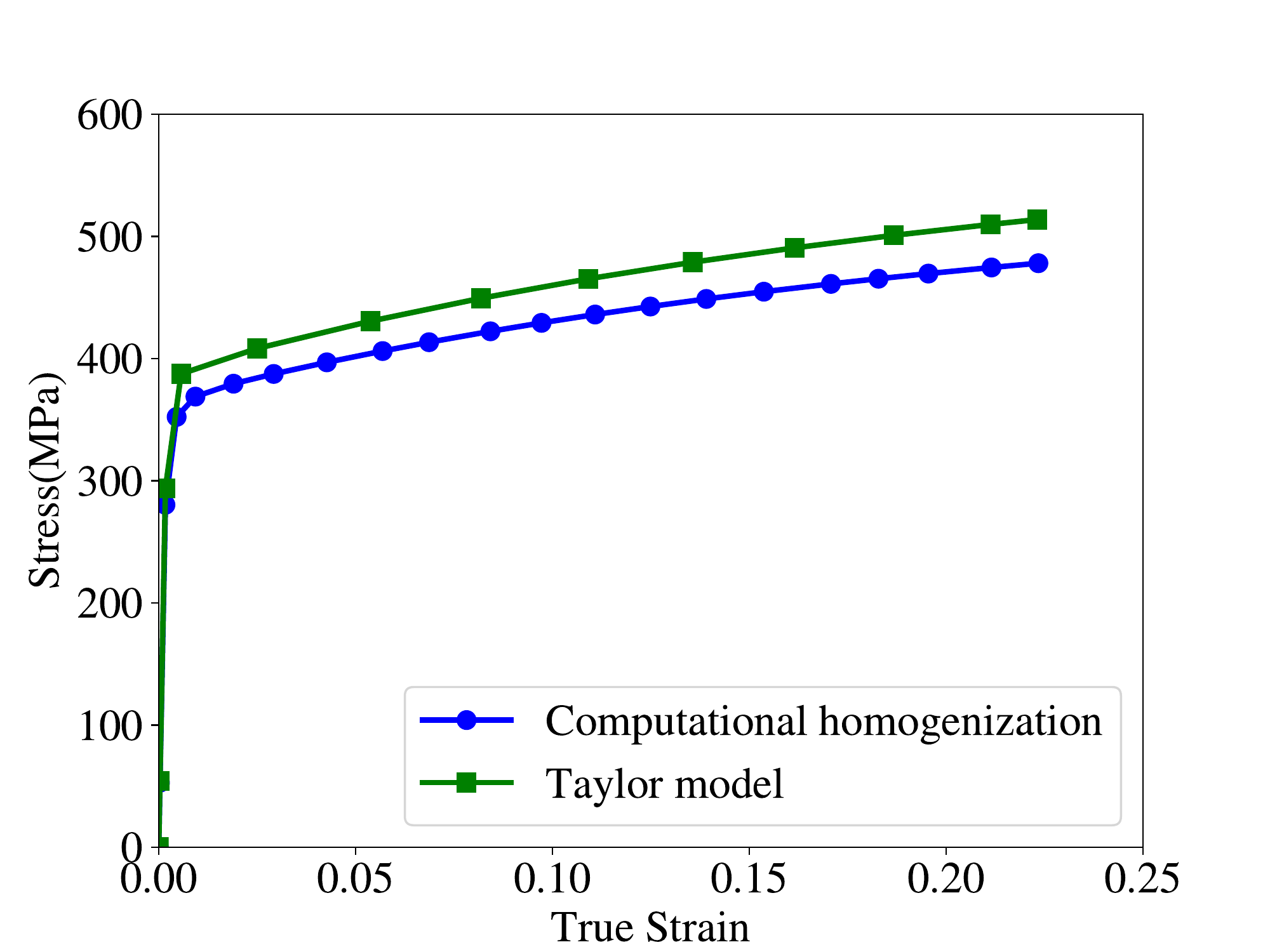}
  }
  \hfill
  \subfigure[]{
    \label{S33ER0005IsoCmp}
    \includegraphics[width=0.45\textwidth]{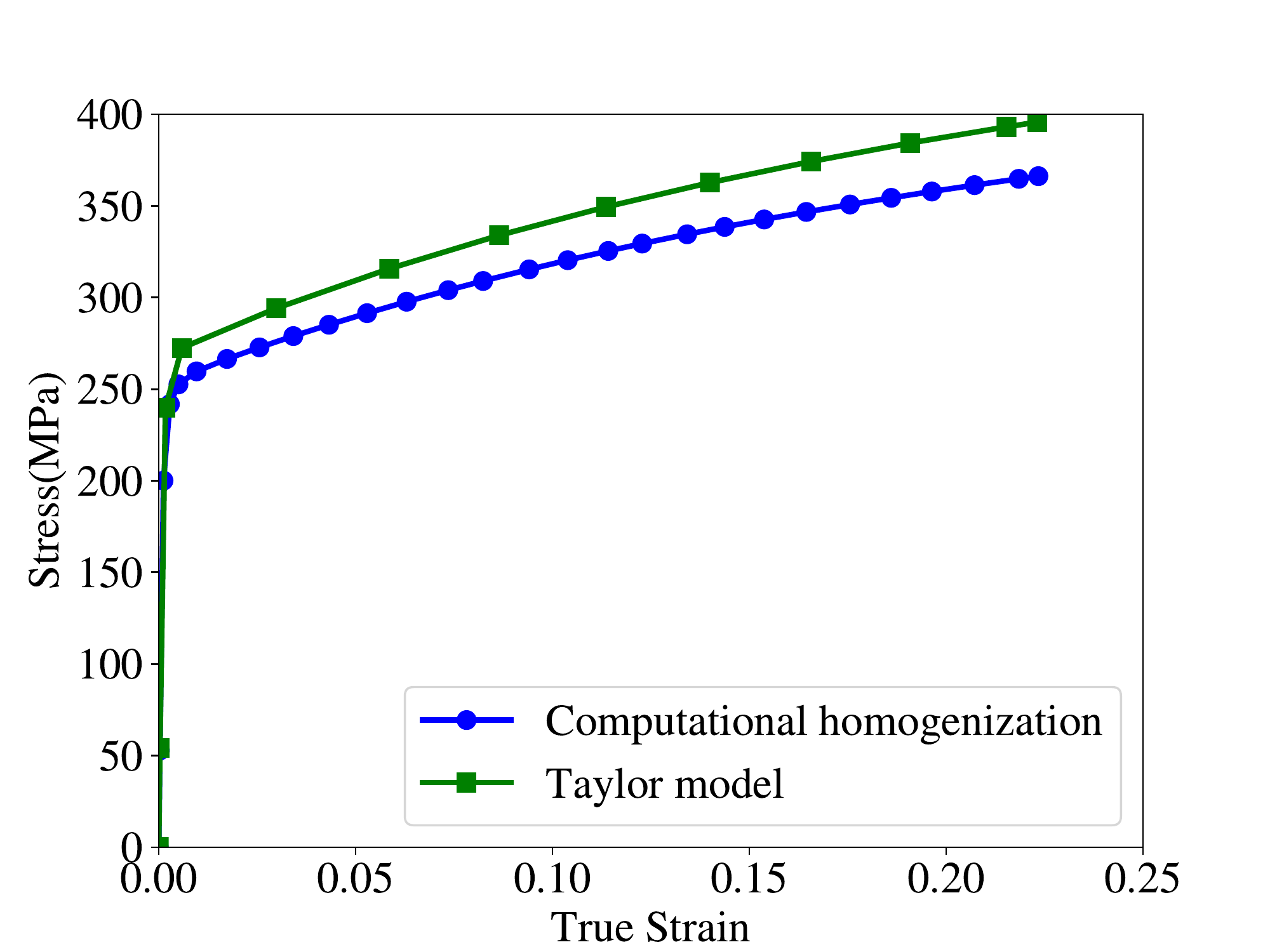}
  }
  \caption{Comparison of stress-strain curves obtained by computational
  homogenization approach and Taylor model of BCC Ta during uniaxial tensile
  tests at various engineering strain rates and isothermal conditions.
   (a). $\dot{\varepsilon}=5.0\cdot10^{3}$ s$^{-1}$
   (b). $\dot{\varepsilon}=5.0$ s$^{-1}$
   (c). $\dot{\varepsilon}=5.0\cdot10^{-3}$ s$^{-1}$}
  \label{Taylor_isothermal}
\end{figure*}

The polycrystal was represented using 400 grains and the orientations of these grains were exactly
the same as those in the RVE. The model parameters were also the same as in previous
simulations. The results under isothermal conditions are represented in
Fig.~\ref{Taylor_isothermal}. It can be observed that, for all the strain rates considered, the
Taylor model produces a stiffer stress-strain response compared with the results obtained by the
computational homogenization approach. The stiffer response of the Taylor model is well known and is
due to the iso-strain assumption. This approach imposes that grains badly oriented for accommodating
the plastic deformation deform identically to well oriented grains, artificially increasing their
elastic strain level and therefore their contribution to the macroscopic stress. It must be noted
that in the isothermal case the temperature does not play any role and the differences between (a),
(b) and (c) are only due to the strain rate sensitivity of the crystal plasticity model.

\begin{figure*}[ht]\centering
  \subfigure[]{
    \label{S33ER5000AdiaCmp}
    \includegraphics[width=0.45\textwidth]{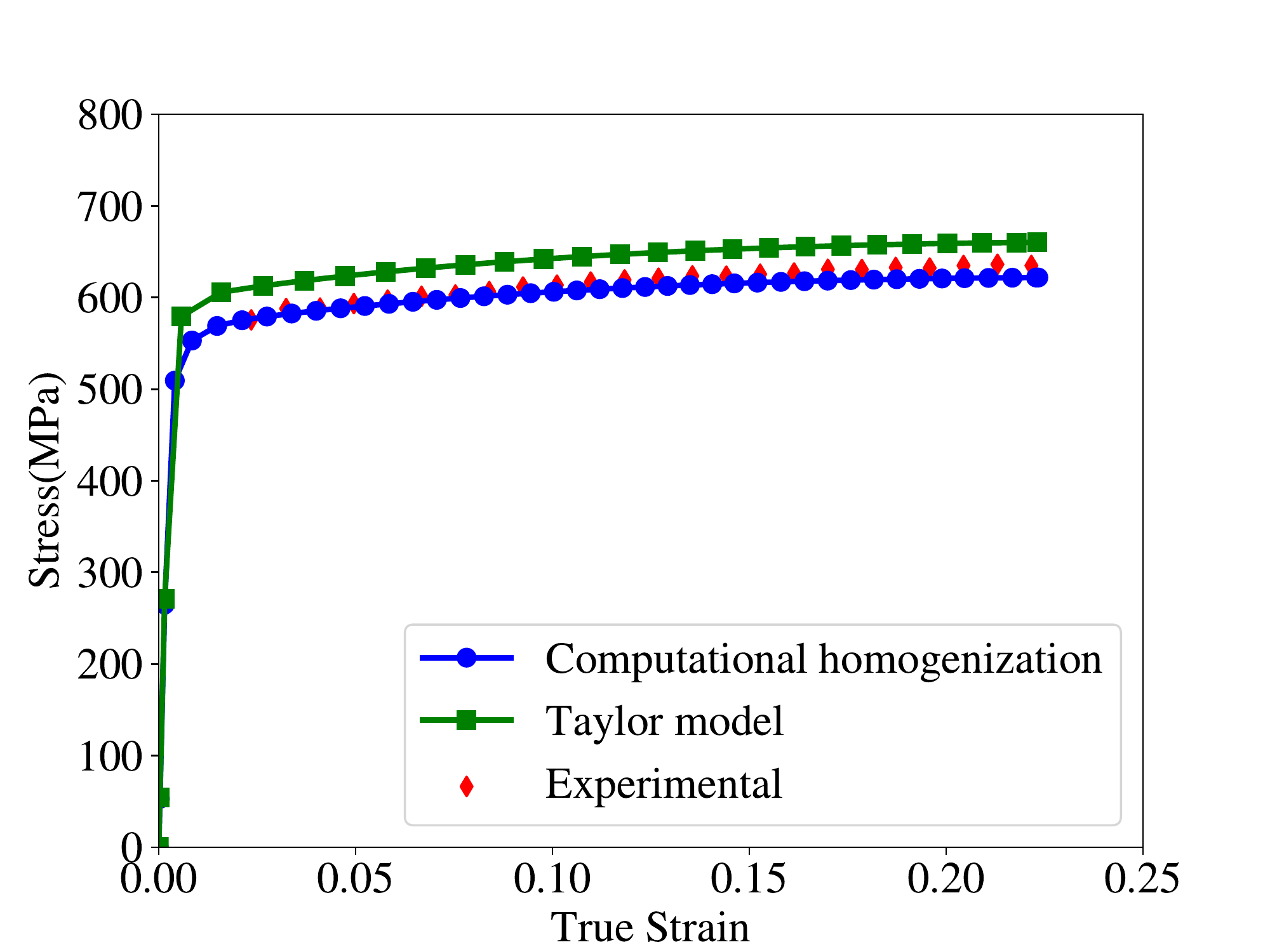}
  }
    \subfigure[]{
    \label{S33ER5AdiaCmp}
    \includegraphics[width=0.45\textwidth]{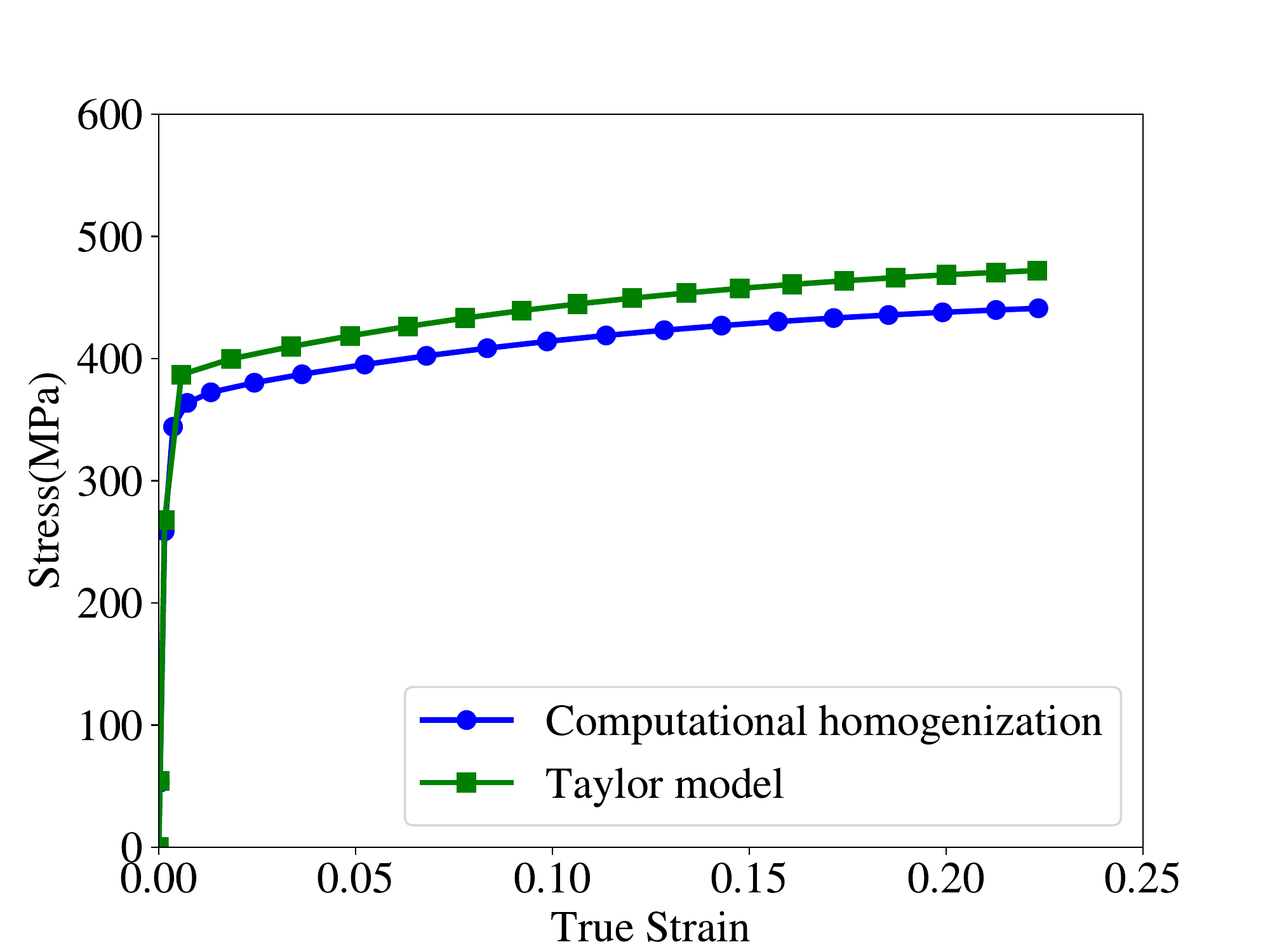}
  }
  \subfigure[]{
    \label{S33ER0005AdiaCmp}
    \includegraphics[width=0.45\textwidth]{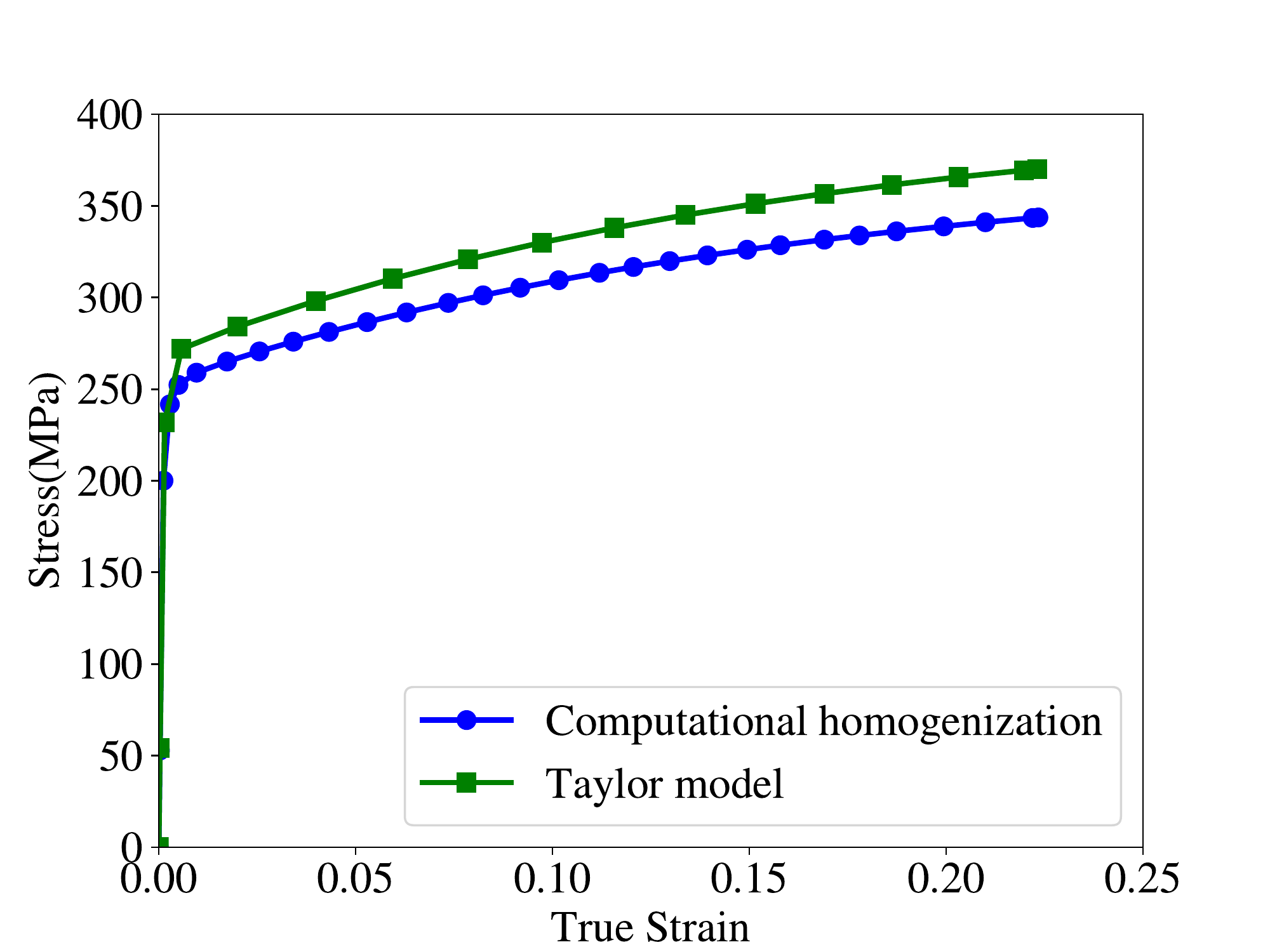}
  }
  \caption{Comparison of stress-strain curves obtained by computational
  homogenization approach and Taylor model of BCC Ta during uniaxial tensile
  tests at various engineering strain rates and adiabatic conditions.
   (a). $\dot{\varepsilon}=5.0\cdot10^{3}$ s$^{-1}$
   (c). $\dot{\varepsilon}=5.0$ s$^{-1}$
   (e). $\dot{\varepsilon}=5.0\cdot10^{-3}$ s$^{-1}$}
  \label{Taylor_adiabatic}
\end{figure*}
The results obtained for macroscopic adiabatic conditions are represented in
Fig.~\ref{Taylor_adiabatic}.  It is shown that, similar to the isothermal case, under an adiabatic
condition Taylor model still gives stiffer stress-strain response compared with the computational
homogenization approach. However, the difference in strength between Taylor model and computational
homogenization for this adiabatic condition is slightly smaller than the difference obtained under
isothermal conditions. This small weakening of the Taylor model respect the computational
homogeneization approach may be due to higher heat generated from overestimated stress response,
which further produces a higher temperature and in return softens the stress response.

More interesting is the comparison of the temperatures obtained using the two homogenization
approaches at the end of the simulation, and listed in Tab. ~\ref{tempDistTaylor}. First, as pointed
out in the previous section, it can be observed that allowing heat diffusion at the microscale
influences the local temperature distribution. This re-distribution of temperature has an effect on
the local behavior and therefore on the actual macroscopic response. In contrast with this effect, it can
be observed that in the Taylor approach temperature differences persist for every strain rate. This
is due to the grain-adiabatic condition used in the Taylor model that does not allow heat
redistribution within the grains, similarly to what happened when microscopic adiabatic condition was
used in the computational model. In summary, the use of a Taylor approach results in a stiffer
response and in a time-independent artificial temperature heterogeneity that might influence the
mechanical results. 

\begin{table}[ht]
  \small
  \centering
  \caption{Comparison of Temperature Distribution in Polycrystal Obtained by Taylor and Computational Homogenization}
  \label{tempDistTaylor}
  \begin{tabular}{ |c|c|c|c|c|c|c| } \hline
    \multirow{2}{*}{}   &  \multicolumn{2}{|c|}{$\dot{\varepsilon}=5000$ s$^{-1}$}   &   \multicolumn{2}{|c|}{$\dot{\varepsilon}=5$ s$^{-1}$} &  \multicolumn{2}{|c|}{$\dot{\varepsilon}=0.005$ s$^{-1}$}   \\
    \cline{2-7}         &   Taylor   &   Comp.    &   Taylor    &    Comp.    &    Taylor    &    Comp.      \\   \hline
    $T_{min}$ (K)       &   340.7    &   316.3    &   327.4     &    333.2    &    319.8     &    324.5      \\   \hline
    $T_{max}$ (K)       &   368.8    &   401.2    &   347.3     &    333.5    &    335.4     &    324.5      \\   \hline
    $T_{ave}$ (K)       &   353.8    &   349.1    &   336.7     &    333.3    &    327.2     &    324.5      \\   \hline
  \end{tabular}
\end{table}

\section{Concluding Remarks}

This work describes a thermo-mechanical crystal plasticity material model and its implementation in a thermo-mechanical finite element framework using strong field coupling and under finite strains. The model is used to obtain the response of a polycrystal by solving the thermo-mechanical problem on a periodic representative volume element of the microstructure.

The most relevant feature of the framework proposed is that, in contrast to phenomenological
macroscopic approaches, the material response is based on the current physical mechanisms
responsible of plastic deformation and is able to relate the macroscopic deformation with the
crystallographic slip at the grain scale, including temperature and rate effects. 
Moreover, the numerical approach is able to resolve the heat diffusion at the microscale introducing
an additional dependency of the macroscopic response with the strain rate, the lower the strain rate
the closer to an isothermal state.

This two-scale approach for simulating thermo-mechanical processes can be used to provide enhanced
predictions and understanding of the material behavior under extreme thermo-mechanical conditions,
both at the macro and the micro scales. The methods presented have been implemented into the
commercial code ABAQUS through user-defined subroutines. The overall scheme has been employed in the
solution of simple numerical examples which illustrate its general features and, in particular, its
ability to capture the main coupling and hardening/softening effects observed in experiments.

\section*{Acknowledgement}

J. Li acknowledges the financial support from China Scholarship Council (Grant
No. 201504890009). I. Romero and J. Segurado have been partially funded by projects DPI2015-67667-C3-1-R and
DPI2015-67667-C3-2-R, respectively.




\end{document}